# Software Development Effort Estimation Using Regression Fuzzy Models

Ali Bou Nassif[1,2,a,*], Mohammad Azzeh[3,b], Ali Idri[4,c] and Alain Abran[5,d]

[1]Department of Electrical and Computer Engineering, University of Sharjah, PO Box 27272, Sharjah, UAE

[2]Department of Electrical and Computer Engineering, University of Western Ontario, London, Ontario, Canada

[3]Department of Software Engineering, Applied Science Private University, PO Box 166, Amman, Jordan

[4]Software Project Management Research Team, ENSIAS, Mohammed V University, Rabat, Morocco

[5]Department of Software Engineering, École de Technologie Supérieure, Montréal, Quebec, Canada

[a]anassif@sharjah.ac.ae, [b]m.y.azzeh@asu.edu.jo, [c]ali.idri@um5.ac.ma, [d]alain.abran@etsmtl.ca

*Corresponding author: Ali Bou Nassif, University of Sharjah, Sharjah, UAE. Email: anassif@sharjah.ac.ae

**Abstract:** Software effort estimation plays a critical role in project management. Erroneous results may lead to overestimating or underestimating effort, which can have catastrophic consequences on project resources. Machine-learning techniques are increasingly popular in the field. Fuzzy logic models, in particular, are widely used to deal with imprecise and inaccurate data. The main goal of this research was to design and compare three different fuzzy logic models for predicting software estimation effort: Mamdani, Sugeno with constant output and Sugeno with linear output. To assist in the design of the fuzzy logic models, we conducted regression analysis, an approach we call 'regression fuzzy logic'. State-of-the-art and unbiased performance evaluation criteria such as standardized accuracy, effect size and mean balanced relative error were used to evaluate the models, as well as statistical tests. Models were trained and tested using industrial projects from the International Software Benchmarking Standards Group (ISBSG) dataset. Results showed that data heteroscedasticity affected model performance. Fuzzy logic models were found to be very sensitive to outliers. We concluded that when regression analysis was used to design the model, the Sugeno fuzzy inference system with linear output outperformed the other models.

## 1. Introduction and Motivation

Generally, estimating project resources continues to be a critical step in project management, including software project development [1]. Ability to predict the cost or effort of a software project has a direct impact on management decision to accept or reject any given project. For example, overestimating software costs may lead to resource wastage and suboptimal delivery time, while underestimation may lead to project understaffing, over budgeting expenses and delayed delivery time [2] [3]. This can lead to loss of contracts and thus, potentially substantial financial losses. Although in practice there is a difference between the expressions, 'software cost estimation' and 'software effort estimation', many authors use either to express the effort required to build a software project measured in person-hours. In this paper, the two expressions are used interchangeably.

Accurate estimation of software resources is very challenging and many techniques have been investigated in order to improve the accuracy of software estimation models [4] [5]. The techniques used in software effort estimation (SEE), are organized into three main groups: expert judgment, algorithmic models and machine learning [6]. Expert judgment depends on the estimator's experience, while algorithmic models use mathematical equations to predict software cost. On the other hand, machine-learning models are based on non-linear characteristics [4]. Algorithmic models and machine-learning models depend on project and cost factors. Among machine-learning models, the fuzzy logic model, first proposed by Lotfi Zadeh [7], has been investigated in the area of software cost estimation by many researchers who have proposed models that outperform the classical SEE techniques [5] [6] [10]. Even so, significant limitations of such models have been identified:

- When examined individually, the performance of different fuzzy logic models seem to fluctuate when tested on different datasets, which can in turn cause confusion around determining the best model [11].

- Most fuzzy logic models were evaluated using mean magnitude of relative error (MMRE), mean magnitude of error relative to the estimate (MMER), relative error (RE) and prediction level (Pred). All these performance evaluation criteria are considered biased [12] [13] [14].
- Several previous studies did not use statistical tests to confirm if the proposed models were statistically different from other models. Failure to employ proper statistical tests would invalidate the results [15].
- Effective design of Sugeno fuzzy logic models with linear outputs, which are scarce in the field of software effort estimation is a challenging task, especially for such models with multiple inputs where identifying the number of input fuzzy sets is in itself challenging.

To address the above limitations, we developed and evaluated three different fuzzy logic models using proper statistical tests and identical datasets extracted from the International Software Benchmarking Standards Group (ISBSG) [16], according to the evaluation criteria proposed by Shepperd and MacDonell [12]. The three models were compared using a multiple linear regression (MLR) and Feed-Forward Artificial Neural Network models developed with the same training and testing datasets used for the fuzzy logic models. This MLR type model was taken to be the base model for SEE.

Among the challenges in designing fuzzy logic models is to determine the number of model inputs and the parameters for the fuzzy Sugeno linear model. To tackle these challenges, we proposed regression fuzzy logic, where regression analysis was used to determine the optimal number of model inputs, as well as the parameters for the fuzzy Sugeno linear model. Note that our regression fuzzy logic (RFL) model should not be confused with fuzzy regression. The latter is actually a regression model that uses fuzzy logic as an input layer [17], whereas RFL is a fuzzy model that uses regression as an input layer. Regarding the fuzzy Sugeno linear model, (***n+1***) parameters are required if the number of inputs is ***n***. MLR models are used to find the (***n+1***) parameters.

In this study, we investigated the following research questions:

RQ1: What is the impact of using regression analysis to tune the parameters of fuzzy models?

To answer this question, we used stepwise regression to determine the number of model inputs and multiple linear regression to adjust the parameters of the Sugeno fuzzy linear model. Then, the three fuzzy logic models, as well as the multiple linear regression model were evaluated using four datasets based on several evaluation performance criteria, such as the mean absolute error, mean balanced relative error, mean inverted balanced relative error, standardized accuracy and the effect size. Statistical tests such as the Wilcoxon test and the Scott-Knott test were used to validate model performance. The mean error of all models was evaluated to determine if the models were overestimating or underestimating.

RQ2: How might data heteroscedasticity affect the performance of such models?

Heteroscedasticity exists as a problem when the variability of project effort increases with projects of the same size. To answer this question, we filtered the ISBSG dataset and divided it into four datasets based on project productivity (effort/size). Homoscedastic datasets are those that have very few variations in project productivity. We studied whether the performance of each model fluctuates when a heteroscedasticity problem exists.

RQ3: How do outliers affect the performance of the models?

To answer this question, we conducted experiments with datasets containing outliers and then repeated the experiments with the outliers removed. We studied the sensitivity of all four models to outliers.

In real life, a machine-learning software estimation model has to be trained on historical datasets. The main objectives of RQ2 and RQ3 are to show that data heteroscedasticity and outliers have a big impact on the performance of the fuzzy-regression estimation models. This would be very helpful in organizations where they have several historical projects. This implies that data cleansing, such as removing outliers and minimizing the data heteroscedasticity effect would be very useful before training the machine-learning prediction model. So identifying these characteristics are of paramount importance, and that this is precisely what best-managed organizations are interested in for estimation purposes. When the software requirements are in such a state of uncertainty, best managed organizations will work first at reducing these uncertainties of product characteristics. For instance, in the medical field, data cleansing is highly important. Causes and effects are identified within a highly specialized context within very specific parameters, and generalization is avoided outside of these selected limitations and constraints.

The contributions of this paper can be summarized as follows:

- To the best of our knowledge, this is the first SEE study that compares the three different fuzzy logic models: Mamdani fuzzy logic, Sugeno fuzzy logic with constant output and Sugeno fuzzy logic with linear output. Both the training and testing datasets were the same for all models. In addition, the three fuzzy logic models were compared to an MLR model. The datasets are from the ISBSG industry dataset. The algorithm provided in Section 4 shows how the dataset was filtered and processed.
- Investigation of the use of regression analysis in determining the number of model inputs, as well as the parameters of the Sugeno model with linear output. We call this approach, 'regression fuzzy'.
- Test the effect of outliers on the performance of fuzzy logic models.
- Investigation of the influence of the heteroscedasticity problem on the performance of fuzzy logic models.

The paper is organized as follows. Section 2 summarizes related work in the field. Section 3 presents additional background information on techniques used in the experiments. The preparation and characteristics of the datasets are defined in Section 4. Section 5 demonstrates how the models were trained and tested. Section 6 discusses the results. Section 7 presents some threats to validity and lastly, Section 8 concludes the paper.

## 2. Related Work

Software effort estimation (SEE) plays a critical role in project management. Erroneous results may lead to overestimating or underestimating effort, which can have catastrophic consequences on project resources. [18]. Many researchers have studied SEE by combining fuzzy logic (FL) with other techniques to develop models that predict effort accurately. Tab1e 1 lists research in FL related to our work.

Table 1 also shows many studies used datasets from the 1970's to the 1990's, such as COCOMO, NASA and COCOMO II, to train and test FL models, and compared performance with linear regression (LR) and COCOMO equations. Moreover, most measured software size as thousands of line of codes (KLOC), several used thousands of delivered source instruction (KDSI) and two used use case points (UCP).

Most studies showed promising results for fuzzy logic (FL) models. Much of the research focus was on Mamdani fuzzy logic models rather than Sugeno fuzzy logic. Only one paper studied the difference between MLR, Mamdani fuzzy logic and Sugeno fuzzy logic with constant parameters [19]. Our study is the first to compare Mamdani to Sugeno with constant output and Sugeno with linear output. The column 'standalone' in Table 1 indicates whether an FL model was used as a standalone model to predict software effort or, alternatively, used in conjunction with other models. In some papers, FL models were compared to neural network (NN), fuzzy neural network (FNN), linear regression (LR) and SEER-SEM models. The evaluation criteria used in related work can be summarized as follows:

- AAE: Average absolute error
- ARE: Average relative error
- AE: Absolute error
- Pred (x): Prediction level
- MMER: Mean magnitude of error relative to the estimate
- MMRE: Mean magnitude of relative error
- VAF: Variance-Accounted-For is the criterion measuring the degree of closeness between estimated and actual value
- RMSE: Root mean squared error
- MdMER: Median magnitude of error relative to the estimate

- MdMRE: Median magnitude of relative error
- ANOVA: Analysis of variance
- RE: Relative error
- MSE: Mean squared error

*Table 1: Related work on fuzzy logic (FL) models for software effort estimation*

| Ref. no | Dataset | | Standalone (yes + type /No) | Comparison conducted | Software Size Unit | Evaluation criteria | Publication year |
|---|---|---|---|---|---|---|---|
| | Source | Size | | | | | |
| 1. [20] | COCOMO'81 | 63 projects | Yes / Sugeno | FL, COCOCMO models | KDSI | AAE, ARE* | 2004 |
| 2. [21] | Artificial + COCOMO81 | 53 projects | Yes/ Mamdani | FL, COCOMO | N&C LOC | AE, Pred(0.25) | 2005 |
| 3. [22] | Private | 41 modules | Yes/ Mamdani | FL, LR | LOC | MMER, Pred(0.20) | 2005 |
| 4. [23] | NASA | 18 projects | Yes/ Sugeno | FL, LR | KLOC | MMRE, VAF**, RMSE | 2006 |
| 5. [6] | Collected by experiment team from 37 developers | 125 projects | Yes/ Mamdani | FL, LR | N&C LOC | MMER, MMRE, Pred(0.25) | 2006 |
| 6. [9] | Collected by experiment team from 37 developers | 125 projects | Yes/ Mamdani | FL (different memberships functions types), LR | N&C LOC | MdMER, Pred(0.25) | 2007 |
| 7. [24] | From source no. 3 & 6 | 10 projects | Yes/ Mamdani | No comparison | LOC | MMRE | 2009 |
| 8. [25] | Private | 200 projects | Yes/ Mamdani | FL, LR | N&C LOC | MMER | 2010 |
| 9. [26] | Artificial + COCOMO81 | - | Yes/ Mamdani | COCOMO/ FL | KDSI | MMRE | 2010 |
| 10. [27] | Private | 24 projects | No/ Sugeno | Use Case Point (UCP) | UCP | MMRE, Pred(0.35) | 2011 |
| 11. [28] | Private | 24 projects | No/ Mamdani | Use Case Point (UCP) | UCP | MMRE, Pred(0.35) | 2011 |
| 12. [29] | COCOMO I, NASA98, data sets, 4 project from software company in Malaysia | 160 projects | No/ Sugeno | FL-COCOMO II / COCOMO II | KSLOC | MMRE, Pred(0.25) | 2011 |
| 13. [30] | Collected by experiment team from 74 developers | 231 projects | Yes/ Mamdani | FL, LR | N&C LOC, reused code | MMER, +ANOVA | 2011 |
| 14. [19] | Collected by experiment team from 37 developers | 125 projects | Yes/ Mamdani + Sugeno_ constant | FL-Mamdani, FL-Sugeno, LR | N&C LOC | MMER, Pred(0.25) | 2013 |
| 15. [31] | COCCOMO NASA | 7 projects | Yes, FL | FL/ NN | LOC | MMRE, Pred(0.25) | 2013 |
| 16. [32] | COCOMO | 69 projects | No | FNN, COCOMO | KESLOC | MMER | 2003 |
| 17. [8] | Artificial | - | No/ Mamdani | COCOMO,81 | KLOC | RE | 2000 |
| 18. [33] | COCOMO'81 | 69 projects | No | FNN, COCOMO | KSLOC | Pred(0.25), MMER | 2007 |
| 19. [34] | COCOMO | 21 | No | FNN, ANN, | KLOC | MMRE, | 2007 |
| | | projects | | COCOMO | | Pred(0.25), MdMRE | |
| 20. [35] | ISBSG Release 9 | 3,024 projects | No | FNN | SLOC | MMRE, MMER, Pred(0.25) | 2009 |
| 21. [36] | NASA | 31 projects | No | FNN /other tools | DKLOC | RMSE, MMRE | 2012 |
| 22. [37] | NASA + Industrial | 99 projects | No | FNN-SEERSEM/ SEERSEM | KLOC | MMRE, Pred(0.3), Pred(0.5), MSE | 2015 |
| 23. [38] | Private | - | No | NN | UCP | MMRE, Pred, MMER | 2012 |

Several limitations are evident in the reported work. First, the majority of the above studies used single datasets for model evaluations. This is a major drawback since the performance of machine-learning models might excel on one dataset and deteriorate on other datasets [39]. Second, most of the models in Table 1 were tested using only MMRE, MMER and Pred (x). Moreover, researchers concentrated on Mamdani-type fuzzy logic and ignored Sugeno fuzzy logic, especially Sugeno with linear output. Furthermore, very few studies used statistical tests to validate their results. Myrveit and Stensrud [15] state that it is invalid to confirm that one model is better than another without using proper statistical tests.

Our paper addressed the above limitations. We developed and compared three different fuzzy logic models using four different datasets. We also used the statistical tests and evaluation criteria proposed by Shepperd and MacDonell [12].

# 3. Background

## *3.1 Fuzzy Logic Model*

In attempting to deal with uncertainty of software cost estimation, many techniques have been studied, yet most fail to deal with incomplete data and impreciseness [40]. Fuzzy logic has been more successful [41] [20]. This is due to the fuzzy nature of fuzzy logic, where model inputs have multiple memberships. Fuzzy logic tends to smoothen the transition from one membership to another [7].

Fuzzy logic (FL) models, generally, are grouped into Mamdani models [42] and Sugeno models [43]. Inputs in FL are partitioned to membership functions with shape types such as: triangular, trapezoidal, bell, etc., which represents how input points are mapped to output [44]. The output of an FL model depends on the model type, i.e. Mamdani or Sugeno. Mamdani FL has its output(s) partitioned to memberships with shapes [45], [46]. On the other hand, in Sugeno models (aka Takagi-Sugeno-Kang model), the output is represented as a linear equation or constant. The Sugeno fuzzy format [43] is given by:

If $f(x_1\ is\ A_1, \dots, x_k\ is\ A_k)$ is the input group, then the output group is $y = g(x_1, \dots, x_k)$ . Thus, the rules are:

If $x_1\ is\ A_1$ and $x_K\ is\ A_k$ then $y = p_0 + p_1 x_1 + \cdots + p_k x_k$

Where k is the number of inputs in the model, and $p_n$ are the coefficients of the linear equation. When the output equation is zero-order, y will be equal to a constant value. In both model types, fuzzy logic has three main parts [47]:

- Fuzzification, which maps the crisp input data to fuzzy sets in order to obtain the degree of equivalent membership.
- Rules, where expert knowledge can be expressed as rules that define the relationship between the input(s) and output.
- Aggregation, which involves firing the rules mentioned above. This occurs by inserting data for the fuzzy model, after which, the resulting shapes from each output are added to generate one fuzzy output.
- De-fuzzification, which involves conversion of the fuzzy output back to numeric output.

### *3.2 Multiple Linear Regression Model*

Regression is one method for representing the relationship between two kinds of variables [48]. The dependent variable, representing the output, is the one that needs to be predicted. The others are called independent variables.. Multiple regression involves many independent variables. A linear relationship between the predicted (dependent) variable and the independent variables can be expressed as follows:

$$Y = \beta_0 + \beta_1 X_1 + \beta_2 X_2 + \cdots + \ +\beta_p X_p + \varepsilon \tag{1}$$

Where $Y$ is the dependent variable, $X_1,\ X_2, \ldots, X_p$ are the independent variables for $p$ number of variables and $\beta_1, \beta_2, \ldots, \beta_p$ are constant coefficients that are produced from the data using different techniques, such as least square error or maximum likelihood, that aim to reduce the error between the approximated and real data. Regardless of technique, error will exist, which is represented by $\varepsilon$ in the above equation.

### *3.3 Evaluation Criteria*

Examining the prediction accuracy of models depends upon the evaluation criteria used. Criteria such as the mean magnitude of relative error (MMRE), the mean magnitude of error relative to the estimate (MMER) and the prediction level (Pred (x)) are well known, but may be influenced by the presence of outliers and become biased [49][12]; therefore, other tests were employed in order to improve the efficiency of the experiments.

- Mean absolute error (MAE) calculates the average of differences in the absolute value between the actual effort (e) and each predicted effort ($\widehat{e}$). The total number of projects is represented as N.

$$MAE_i = \frac{1}{N}\sum_{i=1}^{N} |e_i - \widehat{e_i}| \tag{2}$$

- Standardized accuracy (SA) measures the meaningfulness of model results, which ensures our model is not a random guess. More details can be found in [12].

$$SA = 1 - \frac{MAE}{\overline{MAE_p}} \tag{3}$$

$\overline{MAE_p}$ is the mean value of a large number runs of random guessing.

- Effect size (Δ) tests the likelihood the model predicts the correct values rather than being a chance occurrence.

$$\Delta = \frac{MAE - \ \overline{MAE_p}}{SP_0} \tag{4}$$

Where $SP_0$ is the sample standard deviation of the random guessing strategy.

- Mean balance relative error (MBRE) is given by:

$$\text{MBRE} = \frac{1}{N}\sum_{i=1}^{N} \frac{AE_i}{\min(e_i, \widehat{e_i})} \tag{5}$$

Where $AE_i$ is the absolute error and is calculated as: $AE_i = |(e_i - \widehat{e_i})|$

- Mean inverted balance relative error (MIBRE) is given by:

$$\text{MIBRE} = \frac{1}{N}\sum_{i=1}^{N} \frac{AE_i}{\max(e_i, \widehat{e_i})} \tag{6}$$

- Mean error (ME) is calculated as: $ME = \frac{1}{N}\sum_{i=1}^{N}(e_i - \widehat{e_i}\,)$ (7)

# 4. Datasets

For this research, the ISBSG release 11 [16] dataset was employed to examine the performance of the proposed models. According to Jorgensen and Shepperd [1], utilizing real-life reliable projects in SEE increases the reliability of the study. The dataset contains more than 5,000 industrial projects written in different programming languages and developed using various software development life cycles. Projects are categorized as either a new or enhanced development. Also, the software size of all projects was measured in function points using international standards such as IFPUG, COSMIC, etc. Therefore, to make the research consistent, only projects with IFPUG adjusted function points were considered. The dataset contains more than 100 attributes for each project and includes such items as: project number, project completion date, software size, etc. Also, ISBSG ranks project data quality into four levels, "A" to "D", where "A" indicates projects with the highest quality followed by "B" and so on.

After examining the dataset, we noticed that while some projects had similar software size, effort varied extensively. The ratio between software effort (output) and software size (the main input) is called the productivity ratio. We noticed a substantial difference in the productivity ratio among projects with similar software size. For instance, for the same adjusted function point (AFP), productivity (effort/size) varied from 0.2 to 300. The large difference in productivity ratio makes the dataset heterogeneous. Applying the same model for all projects was therefore not practical. To solve this issue, projects were grouped according to productivity ratio making the datasets more homogeneous. The main dataset was divided into sub datasets, where projects in each sub dataset had only small variations in productivity [50]. For this research, the dataset was divided into three datasets as follows:

- Dataset 1: Small productivity ratio (P), where $0.2 \leq P < 10$;
- Dataset 2: Medium productivity projects where $10 \leq P < 20$; and
- Dataset 3: High productivity ($P \geq 20$).

Also, to evaluate the effect of mixing projects with different productivities together, a fourth dataset was added, which combined all three datasets. Dataset 3 was not as homogeneous as the first two, since productivity in this dataset varied between 20 and 330. This dataset was used to study the influence of data heteroscedasticity on the performance of fuzzy logic models.

Given the ISBSG dataset characteristics discussed above, a set of guidelines for selection of projects was needed to filter the dataset. The attributes chosen for analysis were:

- AFP: adjusted function points, which indicates software size.
- Development type: indicates whether the project is a new development, enhancement or re-development.
- Team size: represents the number of members in each development team.
- Resource level: identifies which group was involved in developing this project such as, development team effort, development support, computer operation support and end users or clients.
- Software effort: the effort in person-hours.

In software effort estimation, it is important to choose non-functional requirements as independent variables, in addition to functional requirements [51]. All of the above features are continuous variables except Resource level which is categorical. The original raw dataset contained 5052 projects. Using the following guidelines to filter the datasets, projects were selected based on:

1. Data quality: only projects with data quality A and B as recommended by ISBSG were selected, which reduced dataset size to 4,474 projects.
2. Software size in function points.
3. Four inputs: AFP, team size, development type and resource level; and one output variable: software effort.

4. New development projects only. Projects that were considered enhancement development, re-development or other types were ignored, bringing the total projects to 1,805.
5. Missing information: Filtering the dataset by deleting all the rows with missing data leaving only 468 fully described projects.
6. Dividing the datasets according to their productivity as explained previously to generate three distinct datasets and a combined one.
7. Dividing each dataset into testing and training datasets by splitting them randomly into 70% / 30%, where: 70% of each dataset was used for training and 30% for testing.

The resulting datasets after applying steps 6 and 7:

a. Dataset 1: with productivity $0.2 \leq P < 10$ consisted of 245 projects with 172 projects for training and 73 projects for testing.
b. Dataset 2: with productivity $10 \leq P < 20$ consisted of 116 projects with 81 projects for training and 35 projects for testing.
c. Dataset 3: with productivity higher than or equal to 20 ($P \geq 20$) consisted of 107 projects with 75 projects for training and 32 projects for testing.
d. Dataset 4: combining projects from all three datasets consisted of 468 projects with 328 projects for training and 140 projects for testing.

Table 2 presents some statistical characteristics of the effort attribute in the four datasets. Before using the dataset, a check is needed as to whether or not the attributes data type can be used directly in the models. As discussed in Section 3, FL models divide the input into partitions to ensure smoothness of transition among input partitions; these inputs should be continuous. If one of the inputs is categorical (nominal), a conversion to a binary input is required [52]. Thus, the resource attribute, a categorical variable, was converted to dummy variables. A further operation was performed on the datasets to remove outliers from the testing dataset. The aim here was to study the effects on the results of statistical and error measurement tests. In other words, we analyzed the datasets with outliers, then without outliers. A discussion of the results is presented in Section 6. Figure 1 shows the boxplot of the four datasets, where stars represent outliers. Datasets 1, 3 and 4 had outliers, while Dataset 2 had none. Removing the outliers from Datasets 1, 3 and 4 reduced their sizes to 65, 29 and 130, respectively; Dataset 2 remained unchanged.

*Table 2: Description of effort attribute in all datasets.*

| *Dataset* | *N* | *Mean* | *StDev* | *Min* | *Max* | *Median* | *Skewness* | *Kurtosis* |
|---|---|---|---|---|---|---|---|---|
| *Effort _ dataset 1* | 245 | 883.6 | 1486 | 12 | 14656 | 397 | 5.23 | 37.17 |
| *Effort _ dataset 2* | 116 | 643 | 887.3 | 31 | 4411 | 280 | 2.28 | 5 |
| *Effort _ dataset 3* | 107 | 367 | 391 | 11 | 2143 | 254 | 2.47 | 6.9 |
| *Effort _ dataset 4* | 468 | 706 | 1194 | 11 | 14656 | 310 | 5.8 | 50.5 |

Note: N: number of projects, StDev: Standard Deviation

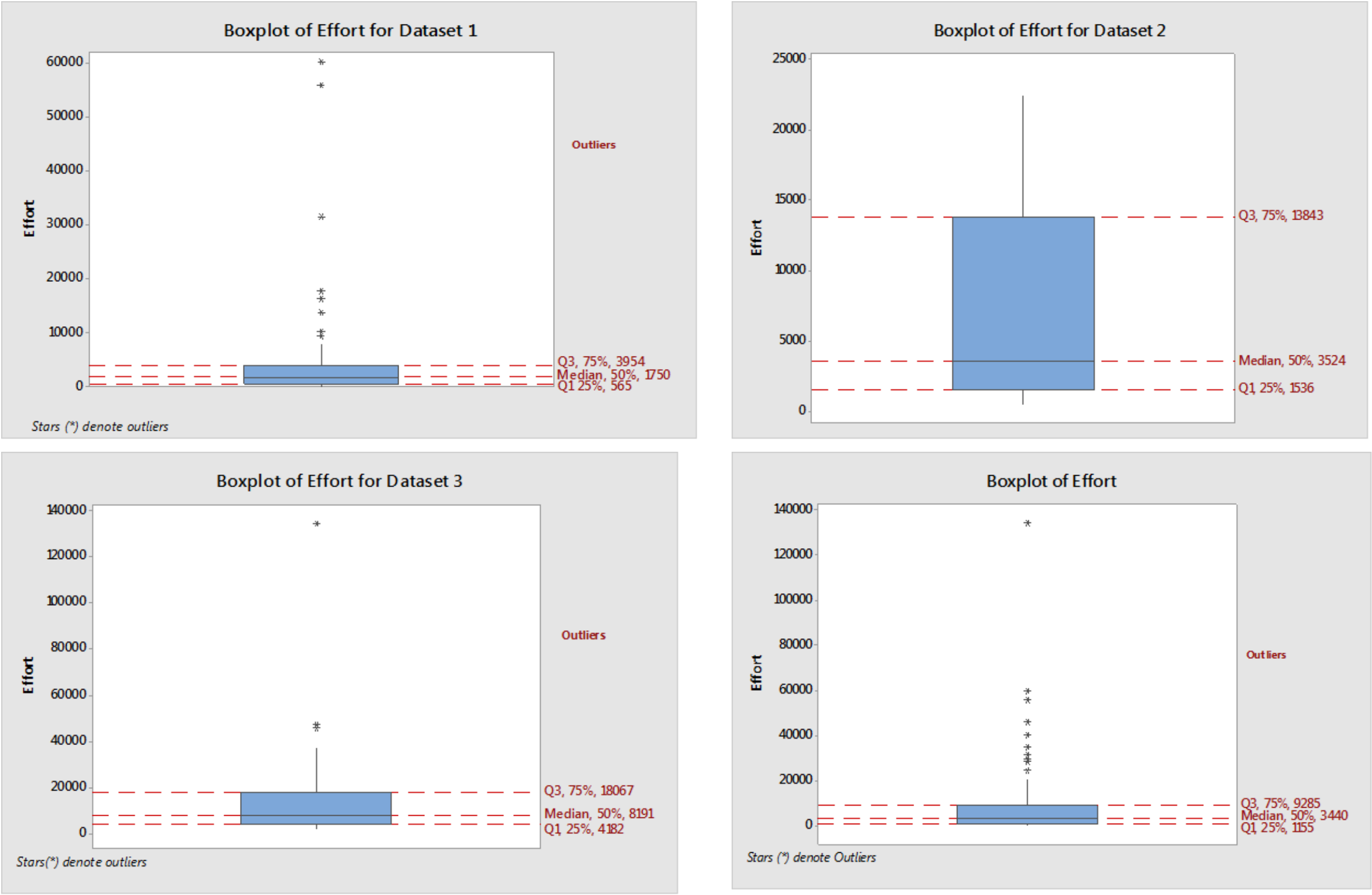

*Figure 1: Boxplot for effort for each dataset*

# 5. Model Design

In this section, the methods used to design the four models, MLR, Sugeno linear FL, Sugeno constant FL and Mamdani FL are presented. The training dataset for each of the four datasets was used to train each model, and then tested using the testing datasets. Performances were analyzed and results presented in Section 6.

As mentioned in Section 4, since all projects have the same development type, the latter was removed as an input, such that three inputs remained for each model. They are: software size (AFP), team size and resource level. The resource level attribute was replaced by dummy variables since it was a categorical variable. A stepwise regression was applied to exclude input variables that were not statistically significant. The same inputs were then utilized for all models in each dataset.

A multiple linear regression model was generated from every training dataset. The fuzzy logic models were then designed using the same input dataset.

To design the Mamdani FL model, the characteristics of each input were examined first, specifically the min, max and average. This gives us a guideline as to the overall shape of memberships. Then, considering that information, all inputs and output were divided into multiple overlapping memberships. Simple rules were written to enable output generation. Usually, simple rules take each input and map it to the output in order to determine the effect of every input on the output. This step can be shortened if some knowledge of the data is available. In our case, since this knowledge existed, setting the rules was expedited. Then, to evaluate and improve the performance of the model, training datasets were randomly divided into multiple sections, and a group tested each time. Rules and memberships were updated depending on the resulting error from those small tests.

Sugeno constant FL has similar characteristics to Mamdani FL, so the same steps were followed except for the output design. The output was divided into multiple constant membership functions. Initial values for each membership function were set by dividing the output range into multiple subsections and then calculating the average of each subsection. Then, the performance of the model was improved by utilizing the training datasets as explained previously.

Lastly, the Sugeno linear FL model was designed. As explained in Section 3, this model is a combination of fuzzy logic and linear regression concepts, each of which is reflected in the design. The steps for designing the input memberships were similar to the steps followed in the Mamdani and Sugeno constant models, whereas the output required a different methodology. The output was divided into multiple memberships, where each membership was represented by a linear regression equation. Hence, the output of the dataset was divided into corresponding multiple overlapping sections, and a regression analysis was applied to each, in order to generate the MLR equation. Subsequently, model performance was improved using the training dataset, as mentioned previously. Note that, over improving the models using training datasets leads to overfitting, where training results are excellent, but testing results are not promising. Therefore, caution should be taken during the training steps. After training, all the models were tested on the testing datasets that were not involved in the training steps.

A summary of the system is shown in Figure. 2.

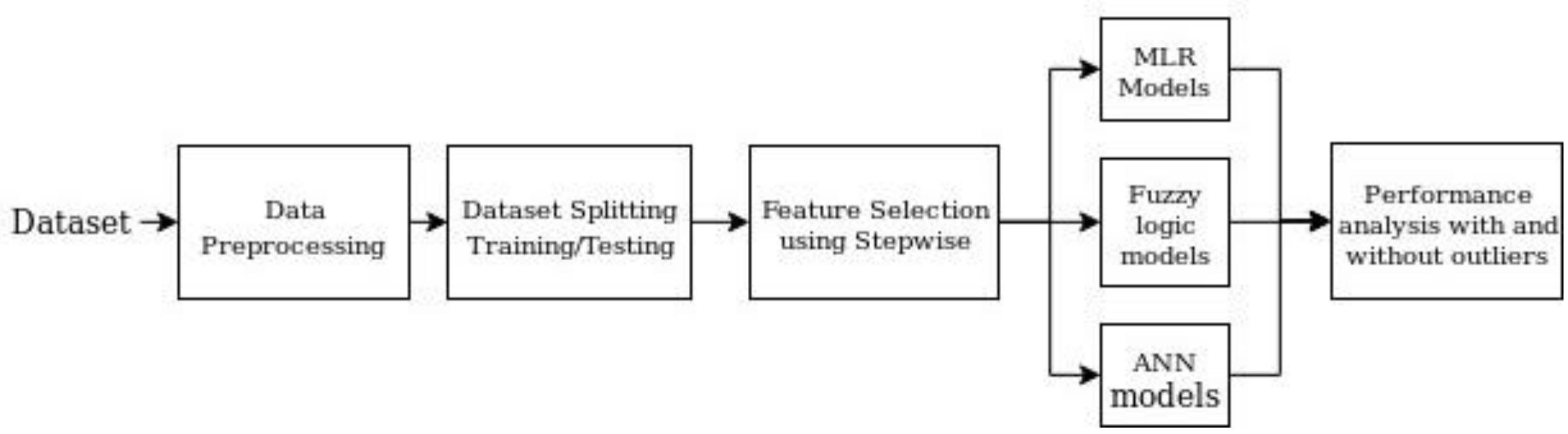

*Figure 2: Block diagram of model design steps*

Table 3a depicts the membership functions (mfs) of the Mamdani, Sugeno constant and Sugeno linear models in the presence of outliers. Tables 3b, 3c and 3d display the parameters of the fuzzy logic models for Dataset 1, Dataset 2 and Dataset 3, respectively. Table 3e displays the parameters of the ANN model

*Table 3a: Fuzzy models memberships*

| Model → / Variable ↓ | Mamdani | | Sugeno constant | | Sugeno linear | | Datasets | | | |
|---|---|---|---|---|---|---|---|---|---|---|
| | # of mf | Type of mf | # of mf | Type of mf | # of mf | Type of mf | Data1 | Data2 | Data3 | Data4 |
| AFP (Input) | 3 | Trimf | 3 | Trimf | 3 | Trimf | Included | Included | Included | Included |
| Team Size (Input) | 3 | Trimf | 3 | Trimf | 3 | Trimf | Included | Included | Included | Included |
| Resource Level (Input) | 1 | Trapmf | 1 | Trapmf | 1 | Trapmf | Included | Excluded | Included | Included |
| Effort (Output) | 3 | Trimf | 3 | Const. | 3 | Linear | Included | Included | Included | Included |

*Table 3b: Parameters of Fuzzy models for Dataset1*

| | Mamdani | Sugeno constant | Sugeno linear |
|---|---|---|---|
| AFP | Small [-350 0 350]<br>Average [140 820 1500]<br>Large [1200 1.5e+04 2e+04] | Small [-350 0 350]<br>Average [140 820 1500]<br>Large [1200 1.5e+04 2e+04] | Small [-350 0 350]<br>Average [140 820 1500]<br>Large [1200 1.5e+04 2e+04] |
| Team Size | Small [-8 0 8]<br>Average [7 20 33]<br>Large [30 50 70] | Small [-8 0 8]<br>Average [7 20 33]<br>Large [30 50 70] | Small [-8 0 8]<br>Average [7 20 33]<br>Large [30 50 70] |
| Resource Level1 | one [0.7 0.7 1 1] | one [0.7 0.7 1 1] | one [0.7 0.7 1 1] |
| Resource Level2 | NA | NA | NA |
| Effort | Small [-2600 0 2600]<br>Average [1500 6000 1.2e+04]<br>Large [9500 5.6e+04 7.84e+04] | Small [973]<br>Average [2882]<br>Large [1.242e+04] | Small [3 116 385 -289]<br>Average [4 278 633 -1332]<br>Large [4.3 361 827 -2013] |

*Table 3c: Parameters of Fuzzy models for Dataset2*

| | Mamdani | Sugeno constant | Sugeno linear |
|---|---|---|---|
| AFP | Small [-260 0 260]<br>Average [200 1450 2700]<br>Large [250 1.5e+04 2e+04] | Small [-260 0 260]<br>Average [200 1450 2700]<br>Large [250 1.5e+04 2e+04] | Small [-260 0 260]<br>Average [200 1450 2700]<br>Large [250 1.5e+04 2e+04] |
| Team Size | Small [-8 0 8]<br>Average [6 15 24]<br>Large [20 100 184] | Small [-8 0 8]<br>Average [6 15 24]<br>Large [20 100 184] | Small [-8 0 8]<br>Average [6 15 24]<br>Large [20 100 184] |
| Resource Level1 | NA | NA | NA |
| Resource Level2 | NA | NA | NA |
| Effort | Small [-3000 0 3000]<br>Average [1000 1e+04 2.2e+04]<br>Large [1e+04 6.5e+04 9.1e+04] | Small [1100]<br>Average [7000]<br>Large [2e+04] | Small [13.56 15.3 -10.4]<br>Average [12.12 135.2 47.7]<br>Large [12.4 115 111] |

*Table 3d: Parameters of Fuzzy models for Dataset3*

| | Mamdani | Sugeno constant | Sugeno linear |
|---|---|---|---|
| AFP | Small [-450 0 450]<br>Average [200 900 1100]<br>Large [892.9 1.5e+04 2e+04] | Small [-450 0 450]<br>Average [200 900 1100]<br>Large [892.9 1.5e+04 2e+04] | Small [-450 0 450]<br>Average [200 900 1100]<br>Large [892.9 1.5e+04 2e+04] |
| Team Size | Small [-8 0 8]<br>Average [5 25 50]<br>Large [35 350 645] | Small [-8 0 8]<br>Average [5 25 50]<br>Large [35 350 645] | Small [-8 0 8]<br>Average [5 25 50]<br>Large [35 350 645] |
| Resource Level1 | One [0.7 0.7 1 1] | One [0.7 0.7 1 1] | One [0.7 0.7 1 1] |
| Resource Level2 | One [0.7 0.7 1 1] | One [0.7 0.7 1 1] | One [0.7 0.7 1 1] |
| Effort | Small [-3000 0 3000]<br>Average [1000 1e+04 2.2e+04]<br>Large [1e+04 6.5e+04 9.1e+04] | Small [4500]<br>Average [1.5e+04]<br>Large [3.48e+04] | Small [34.7 243 -4331 0 2345]<br>Average [22.2 88.4 -1.096e+04 0 1.308e+04]<br>Large [22.23 80.8 -2.042e+04 -2.748e+04 2.45e+04] |

*Table 3e: Parameters of ANN and MLR models for every dataset*

| | ANN (Feed-forward backprop) | MLR |
|---|---|---|
| Dataset 1 | No. of hidden layers: 1<br>No. of hidden neurons: 8 | Y_est=-2674.5+752.9xTeam_Size+1.94xAFP+1413.27x"Resource_Level=1" |
| Dataset 2 | No. of hidden layers: 1<br>No. of hidden neurons : 3 | Y_est= -138.5828+ AFP*12.6030+Team_Size *109.3311 |
| Dataset 3 | No. of hidden layers: 1<br>No. of hidden neurons : 6 | Y_est= 8630.3198+AFP*26.9786+Team_Size*85.1768 +'Resource_Level=1'*-8082.6417+'Resource_Level=2'* -13687.4085 |
| Dataset 4 | No. of hidden layers: 1<br>No. of hidden neurons : 9 | Y_est= 784.5531 + AFP*5.895416+Team_Size*235.3906 + 'Resource_Level=4'*3121.556 |

Regarding the software tools used in this research, MATLAB was used in designing fuzzy logic and neural network models. For statistical tests and analysis, MATLAB, Minitab and Excel have been used. Testing results are analyzed and discussed in Section 6.

# 6. Model Evaluation & Discussion

The following subsections discuss the performance of the models with and without outliers.

## *6.1 Testing Models with Outliers*

The three fuzzy logic models, Sugeno linear, Sugeno constant and Mamdani, were tested on four testing datasets from ISBSG and then compared to the multi-linear regression model. The resulting actual and estimated values were examined using the error criteria: MAE, MBRE, MIBRE, SA and Δ. Table 4 presents the results of the comparisons.

Since MAE measures the absolute error between the estimated and actual value, the model that has the lowest MAE generated more accurate results. As shown in Table 4, Sugeno linear FL generated results – shaded gray- had the lowest MAE among the four datasets. Additional tests using MBRE and MIBRE criteria were also used to examine the accuracy of the data results. The results, as shown in Table 4, indicate that Sugeno linear FL outperformed the other models. Also, SA measures the meaningfulness of the results generated by the models and Δ measures the likelihood that the data was generated by chance. Table 4 shows that the Sugeno linear FL predicted more meaningful results than other techniques across the four datasets. It is also clear from the SA and delta tests that the fuzzy Mamdani model does not predict well when outliers are present, as shown in Table 4.

*Table 4: Error measures and meaningfulness tests*

| | **Dataset 1** | | | | | | **Dataset 2** | | | | | |
|---|---|---|---|---|---|---|---|---|---|---|---|---|
| | **MAE** | **MBRE** | **MIBRE** | **SA** | **Δ** | **ME** | **MAE** | **MBRE** | **MIBRE** | **SA** | **Δ** | **ME** |
| **MLR_out** | 2745.8 | 77 | 220.6 | 61 | 0.3 | 1129.9 | 1418.6 | 26.1 | 19.2 | 80.9 | 0.9 | -910.2 |
| **Fuzzy Lin_out** | 1842.6 | 31.7 | 39.5 | 73.8 | 0.4 | 1225.1 | 1342.9 | 21 | 16.3 | 81.9 | 0.9 | -801.6 |
| **Fuzzy Const_out** | 2779.5 | 244.9 | 45.1 | 60.5 | 0.3 | 1599 | 3674.7 | 85.8 | 40.2 | 50.5 | 0.5 | 2268.4 |
| **Fuzzy Mam_out** | 4118 | 303.2 | 55 | 41.5 | 0.2 | -2454 | 3268.8 | 92.8 | 37.1 | 56 | 0.6 | -2219 |

| | **Dataset 3** | | | | | | **Dataset 4** | | | | | |
|---|---|---|---|---|---|---|---|---|---|---|---|---|
| | **MAE** | **MBRE** | **MIBRE** | **SA** | **Δ** | **ME** | **MAE** | **MBRE** | **MIBRE** | **SA** | **Δ** | **ME** |
| **MLR_out** | 7528.6 | 4.8 | 34.1 | 62.6 | 0.4 | 3696.3 | 5536.3 | 319.2 | 49.7 | 49.6 | 0.3 | 285.5 |
| **Fuzzy Lin_out** | 7241.4 | -96.6 | 32.3 | 64 | 0.4 | 2796.3 | 4925.3 | 176.1 | 60.9 | 55.1 | 0.3 | -58.9 |
| **Fuzzy Const_out** | 8849.9 | 82.1 | 32.2 | 56.1 | 0.4 | 7721.8 | 6646.9 | 413.5 | 57.2 | 39.4 | 0.2 | 1141.4 |
| **Fuzzy Mam_out** | 9332.2 | 76.6 | 37.6 | 53.7 | 0.4 | 2868.6 | 7265.7 | 334.9 | 55.2 | 33.8 | 0.2 | -1759 |

We also examined the tendency of a model to overestimate or underestimate, which was determined by the mean error (ME). ME was calculated by taking the mean of the residuals (difference between actual effort and estimated effort) from each dataset with outliers. As shown in Table 4, all models tended to overestimate in Dataset 3, three models overestimated in dataset 1, and three models underestimated in dataset 2. Surprisingly, Dataset 2 was the only dataset not containing outliers. Nonetheless, the Sugeno linear model outperformed the other models. We then continued to study this problem by repeating the same process after removing the outliers.

To confirm the validity of results, we applied statistical tests to examine the statistical characteristics of the estimated values resulting from the models, as shown in Table 5. We chose the non-parametric Wilcoxon test to check whether each pair of the proposed models is statistically different based on the absolute residuals. The rationale for choosing the non-parametric test was because the absolute residuals were not normally distributed as confirmed by the Anderson-Darling test. The hypothesis tested was:

$H_0$: There is no significant difference between $model_{(i)}$ and $model_{(j)}$

$H_1$: There is a significant difference between $model_{(i)}$ and $model_{(j)}$

If the resulting p-value is greater than 0.05, the null hypothesis cannot be rejected, which indicates that the two models are not statistically different. On the other hand, if the p-value is less than 0.05, then the null hypothesis is rejected. Table 5 reports the results of the Wilcoxon test, with test results below 0.05 shaded in gray. The results of dataset 1 show that Sugeno linear FL was significantly different from all the other models, while for datasets 2 and 4, the Sugeno linear FL & MLR performed similarly, and both were statistically different from Mamdani and Sugeno constant FL. For dataset 3, none of the models performed differently. For this dataset, based on the Wilcoxon test, the models were not statistically different. This is because a heteroscedasticity problem exists in this dataset. The productivity ratio for this dataset (dataset 3) was between 20 and 330 as discussed in Section 4. This huge difference in productivity led to the heteroscedasticity problem and affected the performance of the models.

*Table 5: Wilcoxon test results*

| | **Statistical Test (Dataset 1)** | | | | **Statistical Test (Dataset 2)** | | | |
|---|---|---|---|---|---|---|---|---|
| | **MLR_out** | **Fuzzy Lin_out** | **Fuzzy Const_out** | **Fuzzy Mam_out** | **MLR_out** | **Fuzzy Lin_out** | **Fuzzy Const_out** | **Fuzzy Mam_out** |
| **MLR_out** | X | 0.002824 | 0.567709 | 0.007086 | X | 0.510679 | 0.012352 | 0.093017 |
| **Fuzzy Lin_out** | 0.002824 | X | 0.007004 | 1.94E-06 | 0.510679 | X | 0.005372 | 0.024118 |
| **Fuzzy Const_out** | 0.567709 | 0.007004 | X | 0.001765 | 0.012352 | 0.005372 | X | 0.646882 |
| **Fuzzy Mam_out** | 0.007086 | 1.94E-06 | 0.001765 | X | 0.093017 | 0.024118 | 0.646882 | X |

| | **Statistical Test (Dataset 3)** | | | | **Statistical Test (Dataset 4)** | | | |
|---|---|---|---|---|---|---|---|---|
| | **MLR_out** | **Fuzzy Lin_out** | **Fuzzy Const_out** | **Fuzzy Mam_out** | **MLR_out** | **Fuzzy Lin_out** | **Fuzzy Const_out** | **Fuzzy Mam_out** |
| **MLR_out** | X | 0.877285 | 0.456147 | 0.643195 | X | 0.373822 | 0.004692 | 0.024525 |
| **Fuzzy Lin_out** | 0.877285 | X | 0.456147 | 0.464303 | 0.373822 | X | 0.000591 | 0.003788 |
| **Fuzzy Const_out** | 0.456147 | 0.456147 | X | 0.177199 | 0.004692 | 0.000591 | X | 0.588519 |
| **Fuzzy Mam_out** | 0.643195 | 0.464303 | 0.177199 | X | 0.024525 | 0.003788 | 0.588519 | X |

One of the tests used to examine the stability of the models was the Scott-Knott test, which clusters the models into groups based on data results using multiple comparisons in one way ANOVA [53]. Models were grouped without overlapping, i.e. without classifying one model into more than one group. Results were obtained, simply, from the graphs.

The Scott-Knott test uses the normally distributed absolute error values of the compared models. Therefore, if the values are not normally distributed, a transformation should take place using the Box-Cox algorithm [54], which was the case in our study.

The models to be compared are lined along the x-axis sorted according to rank, with transformed mean error showing across the y-axis. The farther a model from the y-axis is, the higher the rank is. The vertical lines indicate the statistical results for each model. Models grouped together have the same color. The mean of transformed absolute error is shown as a circle in the dashed line. The results of Scott-Knott tests are shown in Figure 3. The Sugeno linear model was grouped alone in dataset 1 and, was also the highest rank in datasets 1, 2 and 4. In dataset 3, where there was a heteroscedasticity issue, the models showed similar behavior. Nevertheless, the Sugeno linear model was among the highest ranked. MLR was ranked second twice and third twice, generally showing stable average performance, while the other FL models did not show stable behavior. This demonstrates that the Sugeno linear model was stable and provides higher accuracy.

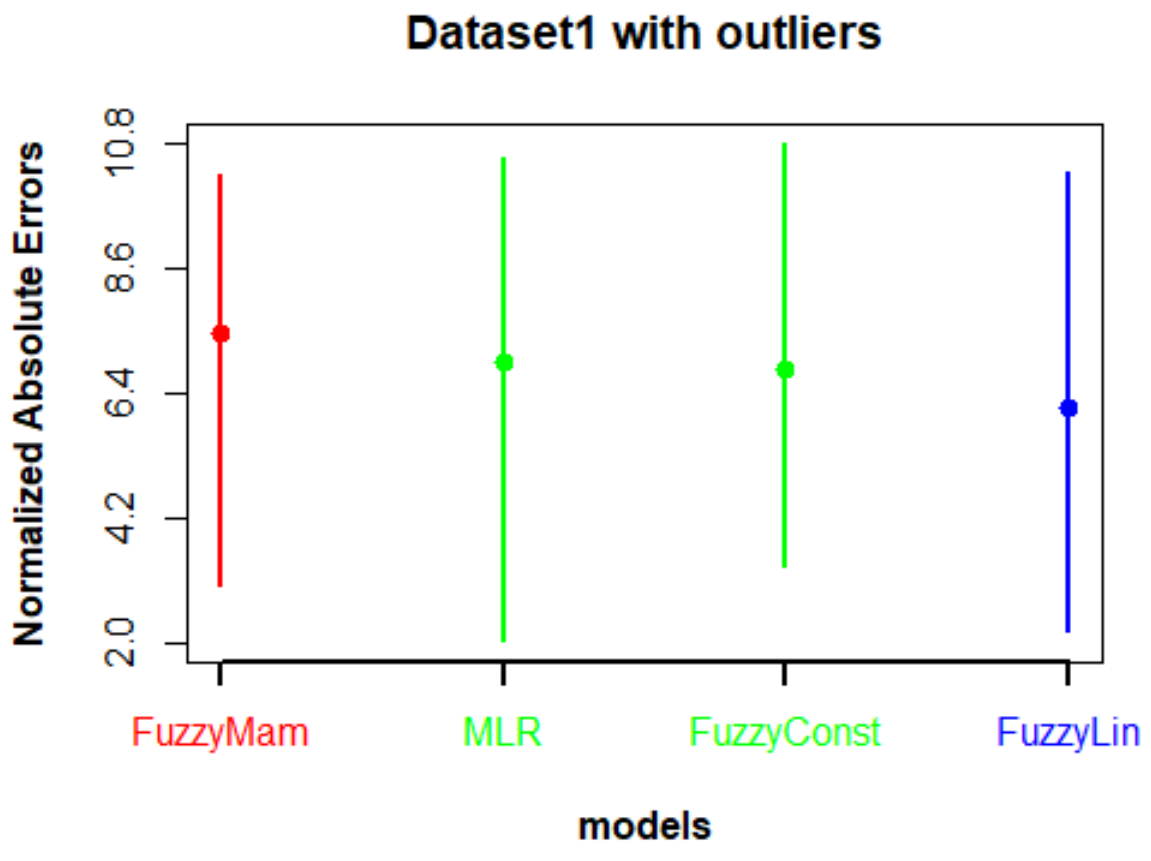

(a) Dataset 1

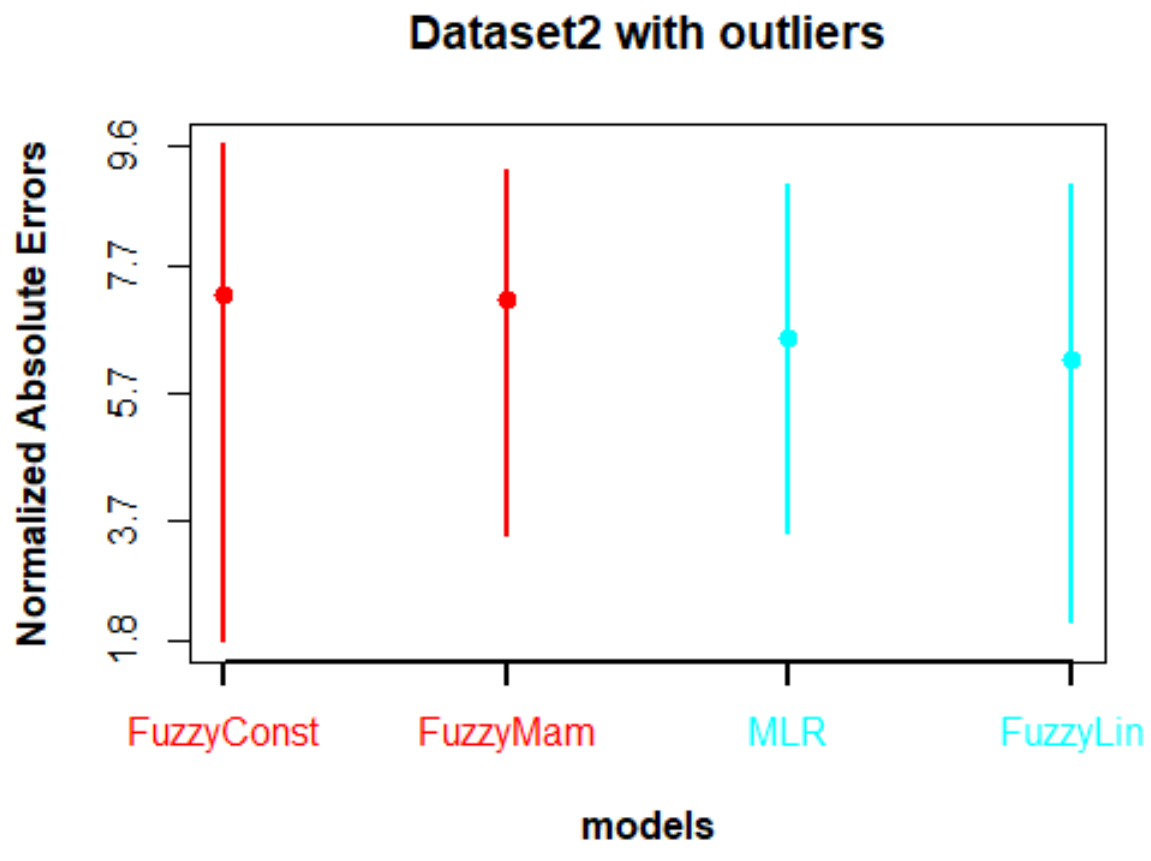

(b) Dataset 2

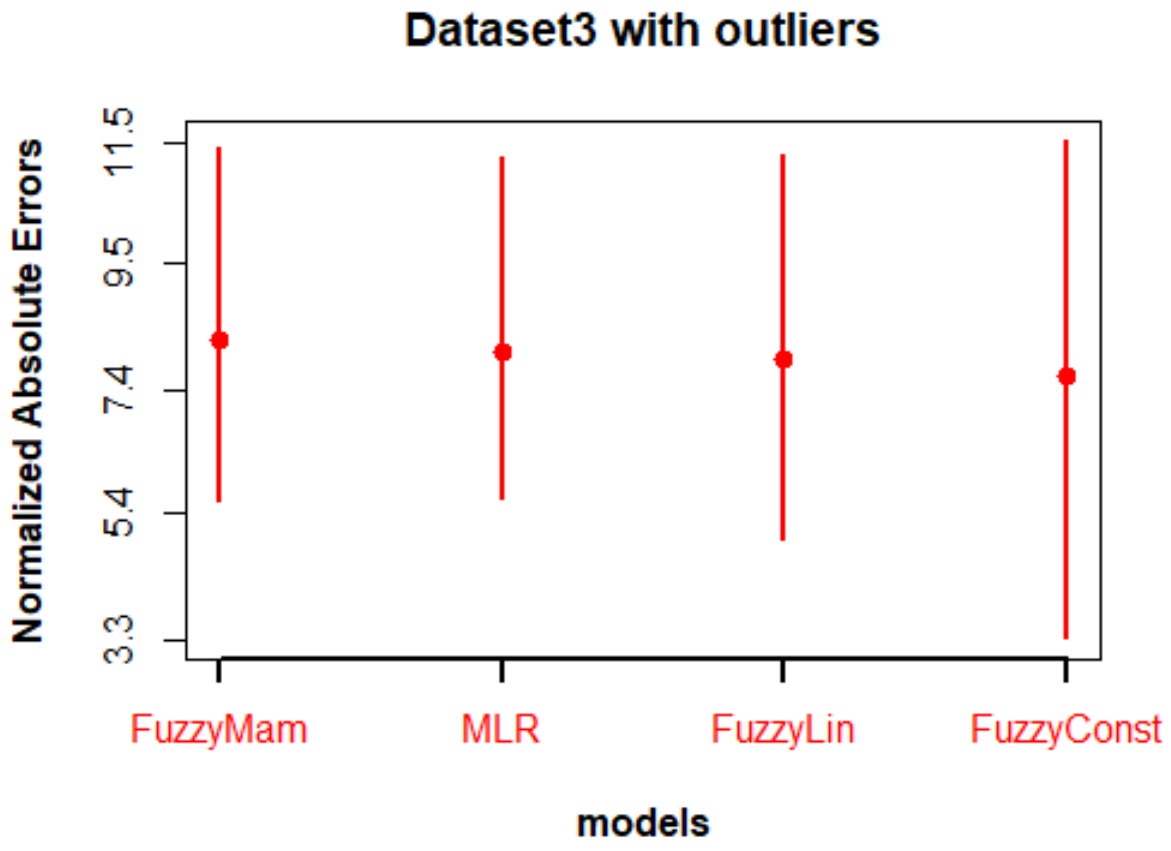

(c) Dataset 3

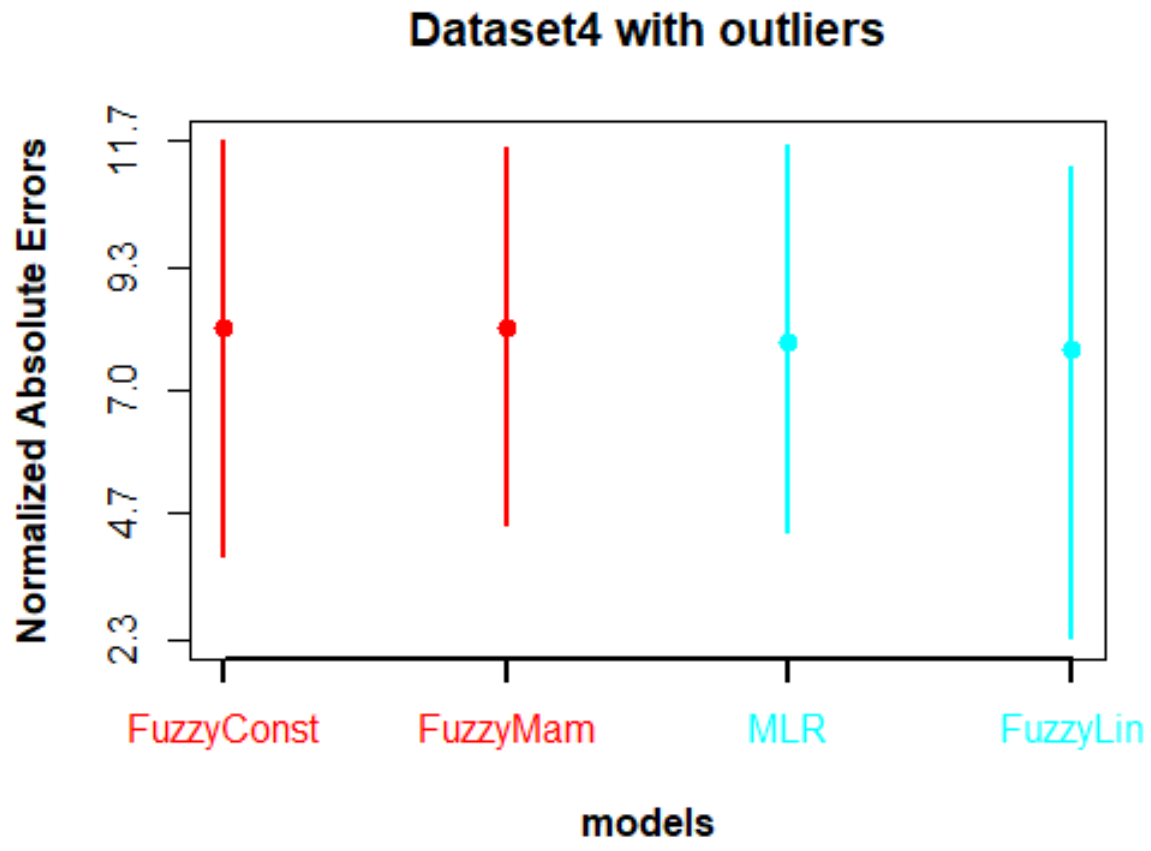

(d) Dataset 4

*Figure 3: Scott-Knott test results in datasets with outliers*

## *6.2 Testing Models without Outliers*

In this section, the models were examined again to study the effect of outliers on model performance. The outliers were removed from the four datasets and the same statistical tests and error measurement tools were applied to the generated results. The filtered datasets were then used for testing the models. We used the interquantile range (IQR) method to determine the outliers. The IQR is defined as: IQR = Q3 – Q1 where Q3 and Q1 are the upper and lower quantile, respectively. Any object that is greater than Q3 + 1.5 IQR or less than Q1 – 1.5 IQR was considered an outlier, since the region between Q1 – 1.5 IQR and Q3 + 1.5 IQR contains 99.3% of the objects [55].

An Interval plot for mean absolute error was generated for all the models using the four testing datasets with and without outliers as depicted in Figure 4. Since the interval plot was for MAE results, the closer the midpoint of each variable to zero, the better it performed. Also, the shorter the interval range the better and more accurate the results. Therefore, it can be concluded from the plots that the general behavior of all the models was improved after removing the outliers. The results were more accurate and the range interval decreased, while the midpoint was closer to zero. The Sugeno linear FL model was markedly more accurate than the other models with or without outliers. It is fair to note that the MLR model had equivalent behavior to the Sugeno linear FL in dataset 2.

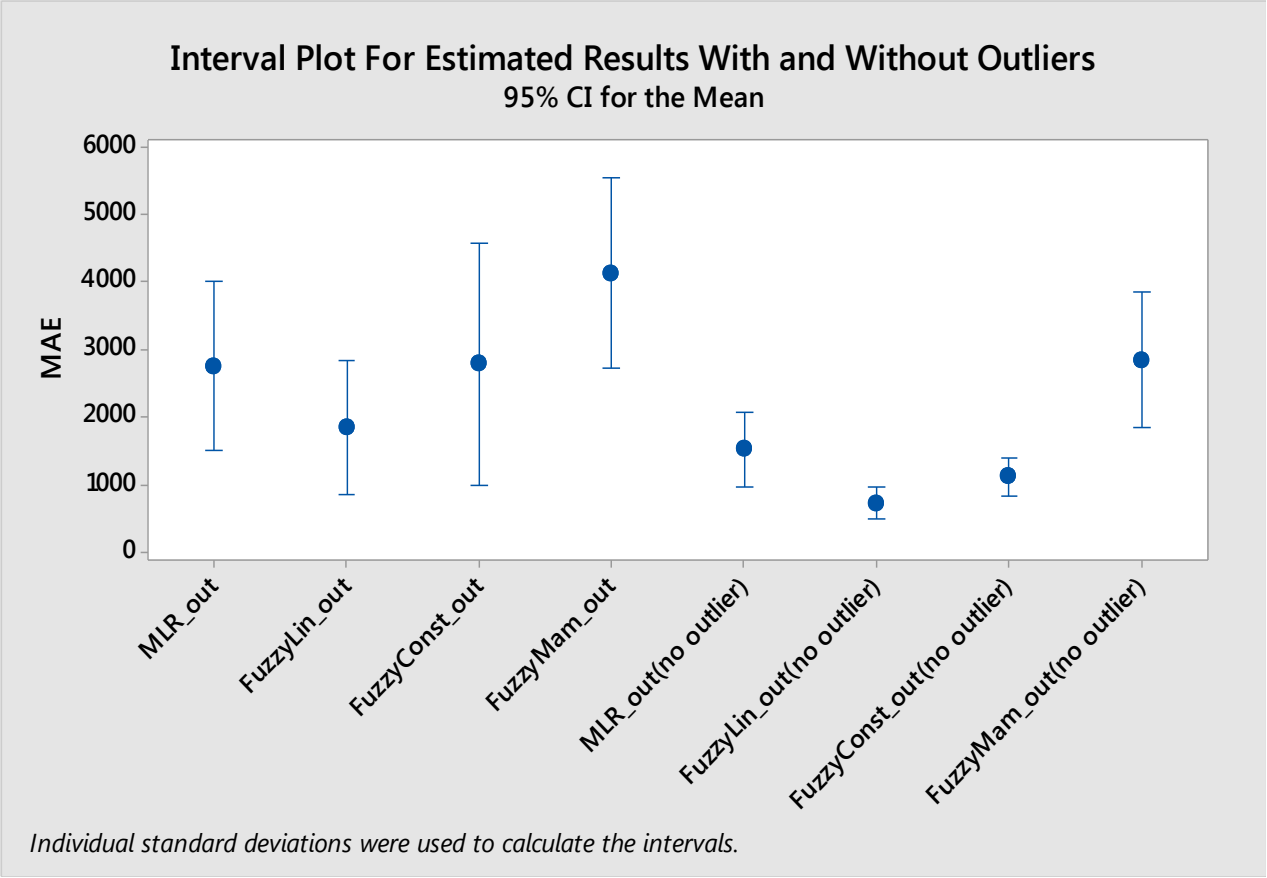

(a) Dataset 1

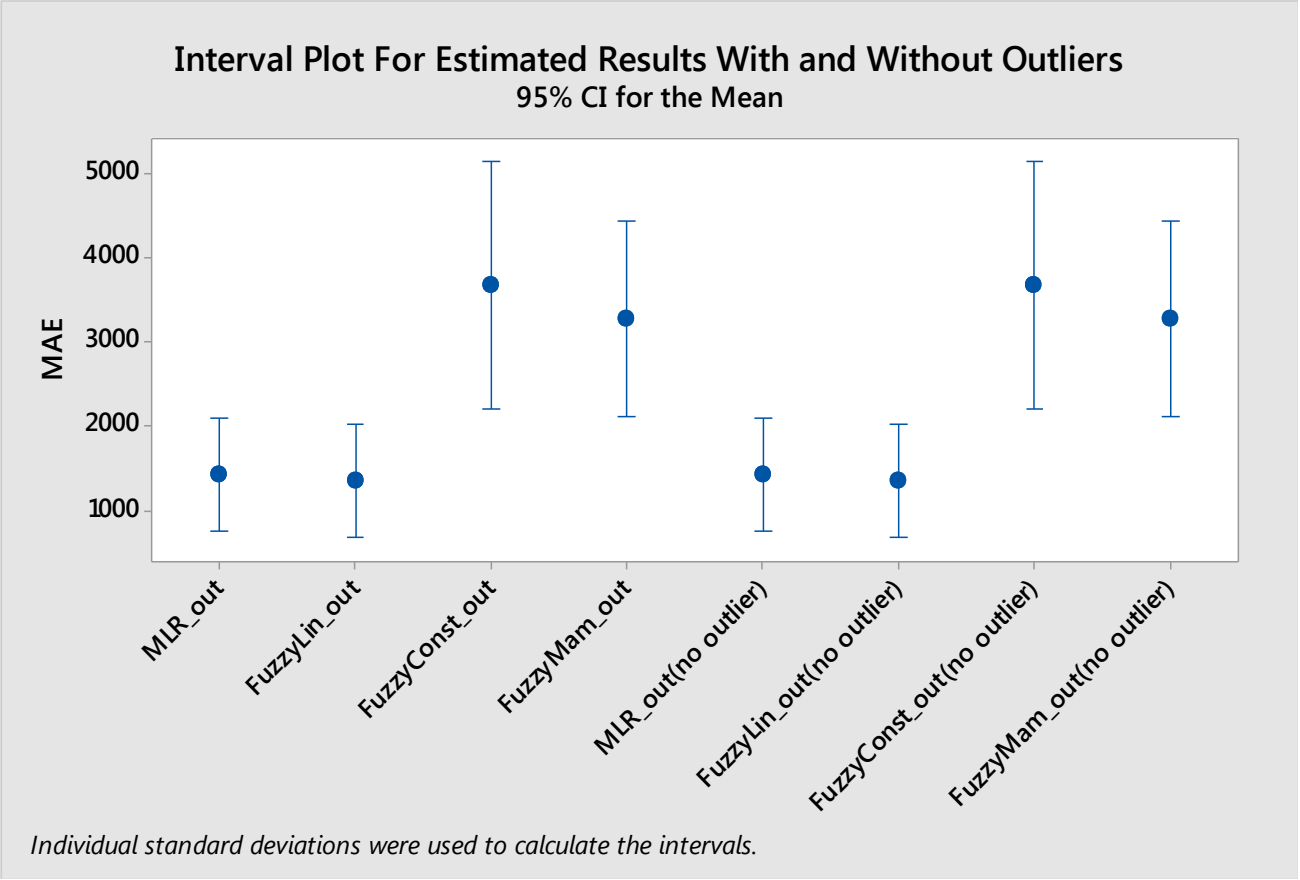

(b) Dataset 2

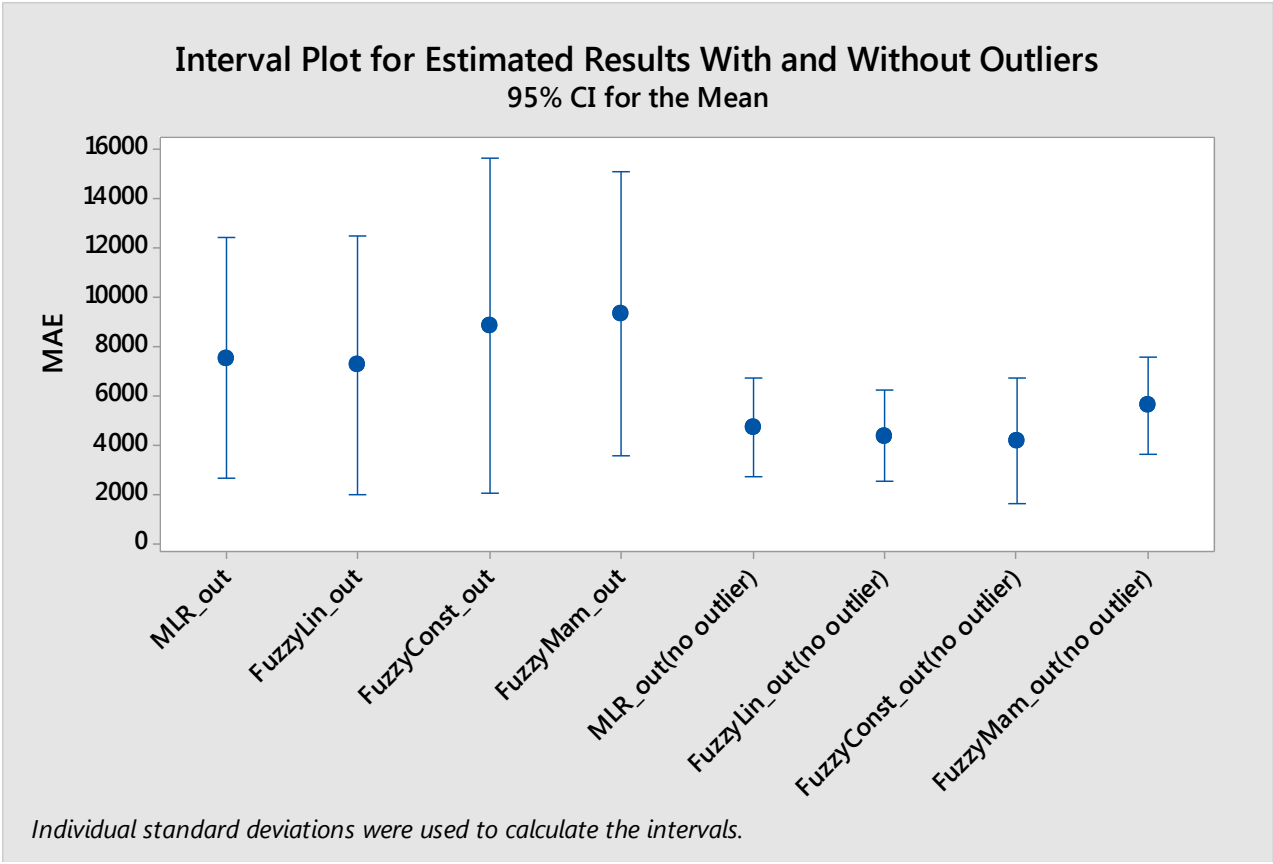

(c) Dataset 3

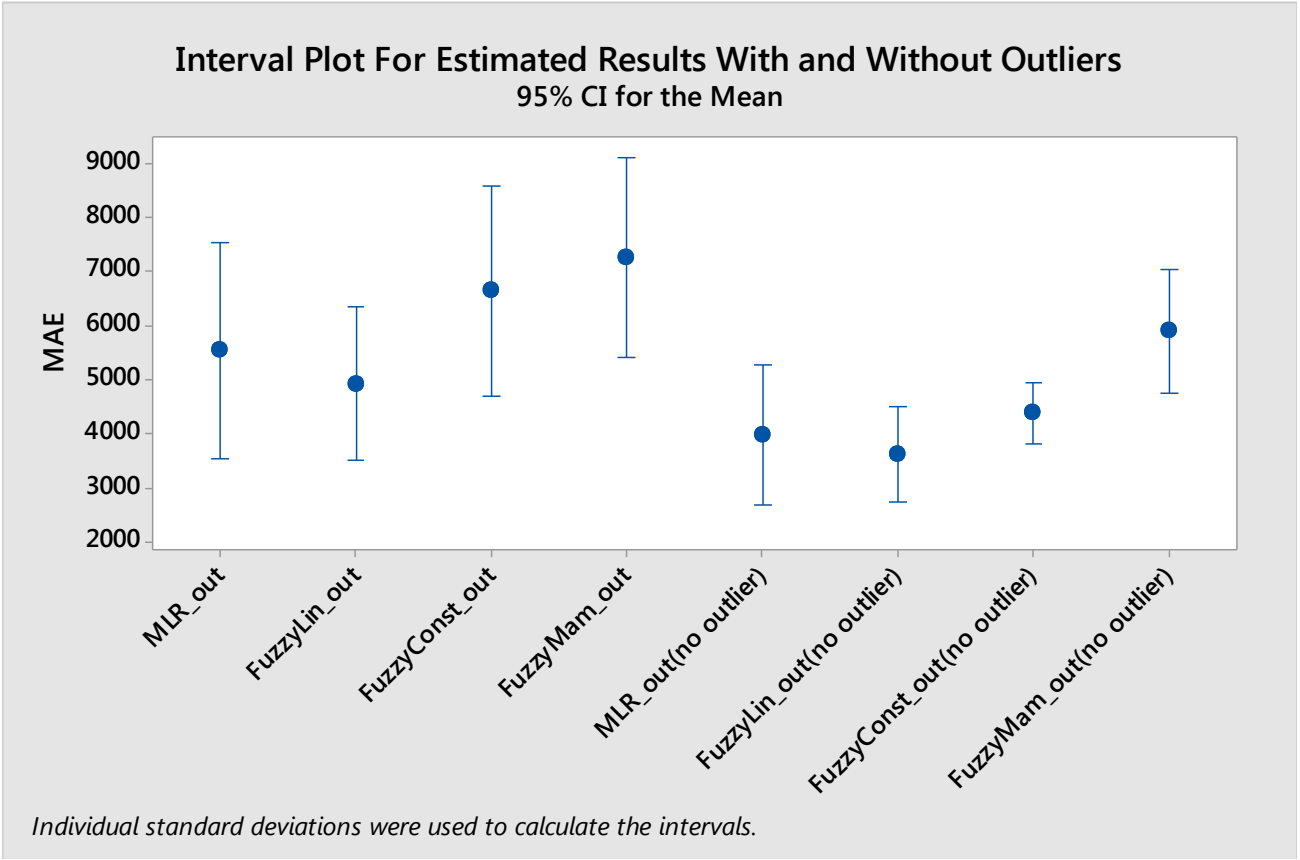

(d) Dataset 4

*Figure 4: Interval plots for estimated results with and without outliers*

To examine the improvement resulting from removal of the outliers, the same error measures were applied to datasets without outliers. Table 6 presents the results for MAE, MBRE, MIBRE, SA and Δ.

*Table 6: Error measures and meaningfulness tests for datasets without outliers*

| | **Dataset 1** | | | | | | **Dataset 2** | | | | | |
|---|---|---|---|---|---|---|---|---|---|---|---|---|
| | **MAE** | **MBRE** | **MIBRE** | **SA** | **Δ** | **ME** | **MAE** | **MBRE** | **MIBRE** | **SA** | **Δ** | **ME** |
| **MLR_out** | 1518.4 | 72.4 | 241.7 | 36.1 | 0.3 | -296.5 | 1418.6 | 26.1 | 19.2 | 80.9 | 0.9 | -910.2 |
| **Fuzzy Lin_out** | 720. | 26.5 | 39.3 | 69.7 | 0.6 | 26.6 | 1342.9 | 21 | 16.3 | 81.9 | 0.9 | -801.6 |
| **Fuzzy Const_out** | 1111.3 | 255.6 | 44.8 | 53.2 | 0.4 | -214.5 | 3674.7 | 85.8 | 40.2 | 50.5 | 0.5 | 2268.4 |
| **Fuzzy Mam_out** | 2834 | 330.1 | 56.6 | -19.2 | 0.2 | -2774.5 | 3268.8 | 92.8 | 37.1 | 56 | 0.6 | -2219 |
| | | | | | | | | | | | | |
| | **Dataset 3** | | | | | | **Dataset 4** | | | | | |
| | **MAE** | **MBRE** | **MIBRE** | **SA** | **Δ** | **ME** | **MAE** | **MBRE** | **MIBRE** | **SA** | **Δ** | **ME** |
| **MLR_out** | 4742.1 | -2.2 | 33.6 | 53.2 | 0.5 | 513.4 | 3982 | 333.7 | 50 | 32.2 | 0.3 | -1673 |
| **Fuzzy Lin_out** | 4376.3 | -114.9 | 31.9 | 56.8 | 0.6 | -528.6 | 3613.7 | 181.8 | 62.5 | 38.5 | 0.4 | -1287 |
| **Fuzzy Const_out** | 4187.5 | 66.7 | 28.7 | 58.7 | 0.6 | 2891.3 | 4377.7 | 421.5 | 56.1 | 25.4 | 0.3 | -1551 |
| **Fuzzy Mam_out** | 5608.5 | 70.7 | 35.8 | 44.7 | 0.5 | -1523.9 | 5897.6 | 348.2 | 55.9 | -0.4 | 0 | -3807 |

Note: MAE stands for mean absolute error, SA stands for standardized, Δ (delta) represents effect size, MBRE stands for mean balance relative, MIBRE stands for mean inverted balance relative error.

Finally, the mean error (ME) from each dataset was calculated to check the effect of removing outliers on overestimating and underestimating project effort. We noticed that the majority of models tend to underestimate after removing the outliers. This confirms the findings of the test on the datasets with outliers, where models tended to overestimate.

The performance of all models without outliers was improved, as the data in Table 6 indicates. We conclude that FL models are sensitive to outliers.

In addition, we examined the effect of outlier removal using the Scott-Knott test. Figure 5 shows the results of the Scott-Knott test. Generally, our conclusions about model stability did not change. However we noted that the mean of transformed absolute error decreased. This shows that removing the outliers increases the accuracy of the models. We conclude that the Sugeno linear FL model was the superior model, both in the presence and absence of outliers.

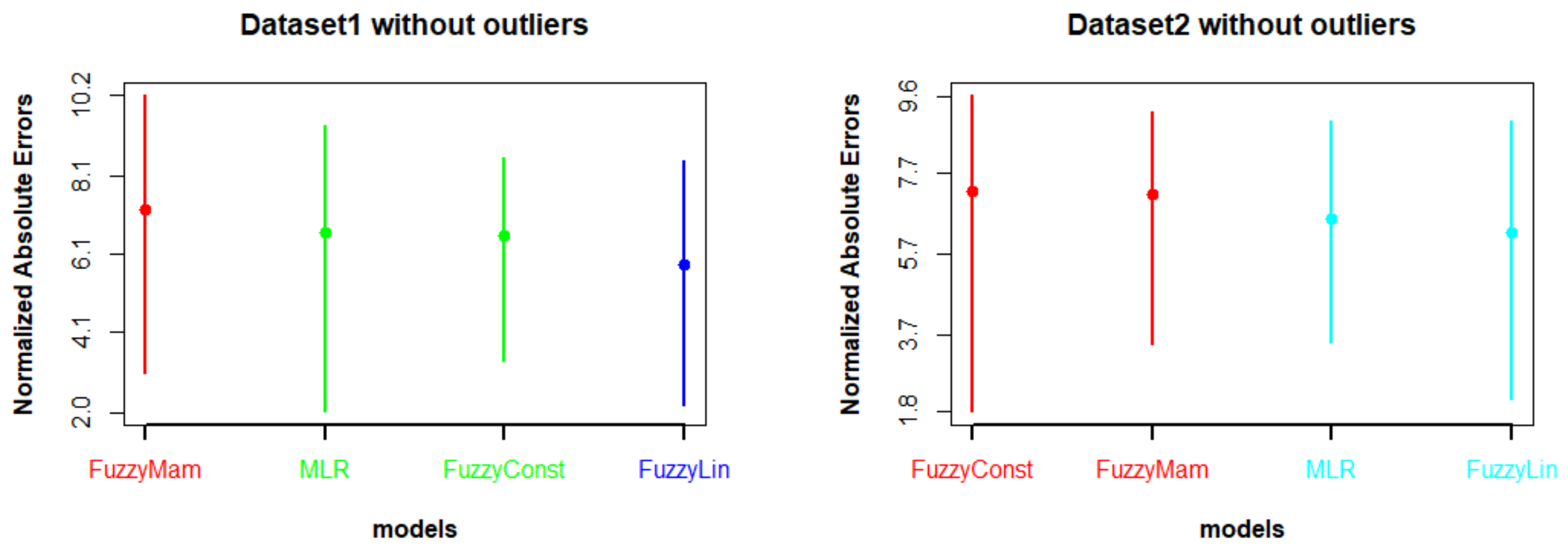

(a) Dataset 1 (b) Dataset 2

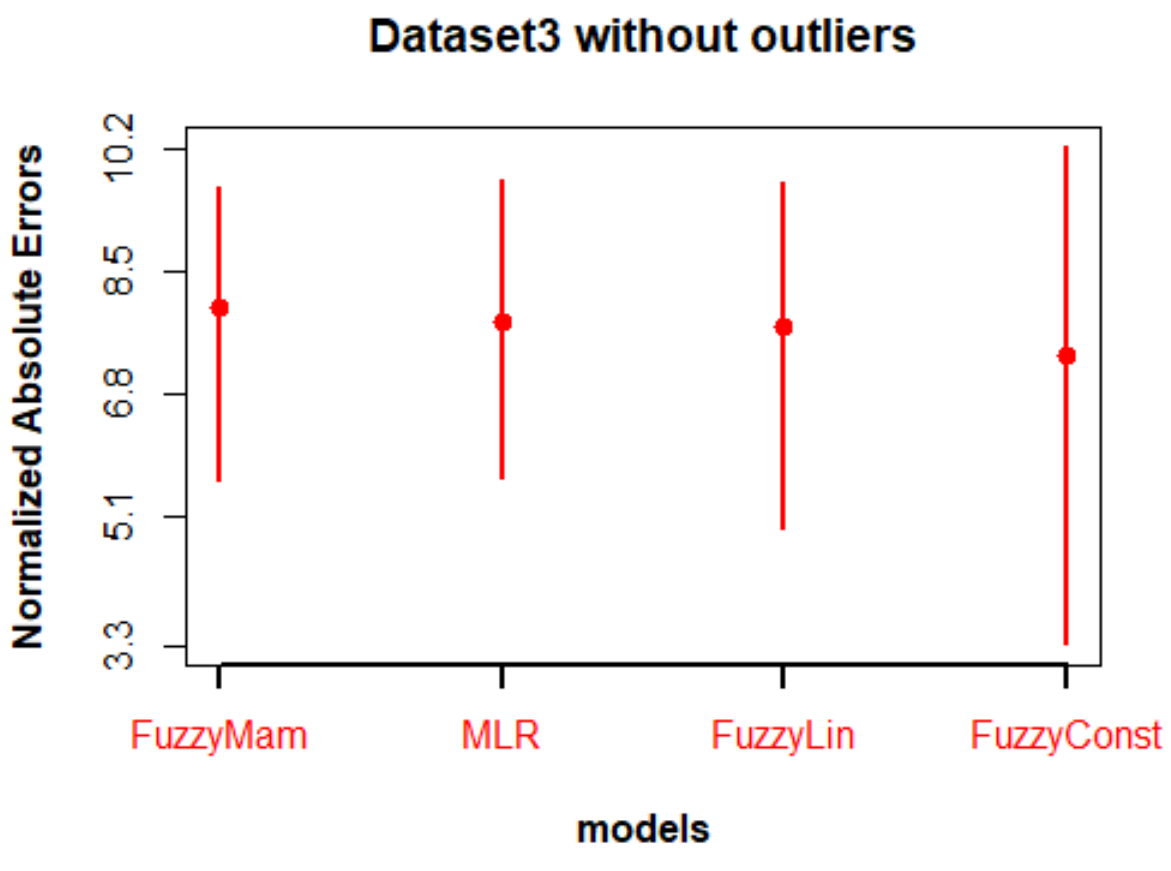

(c) Dataset 3

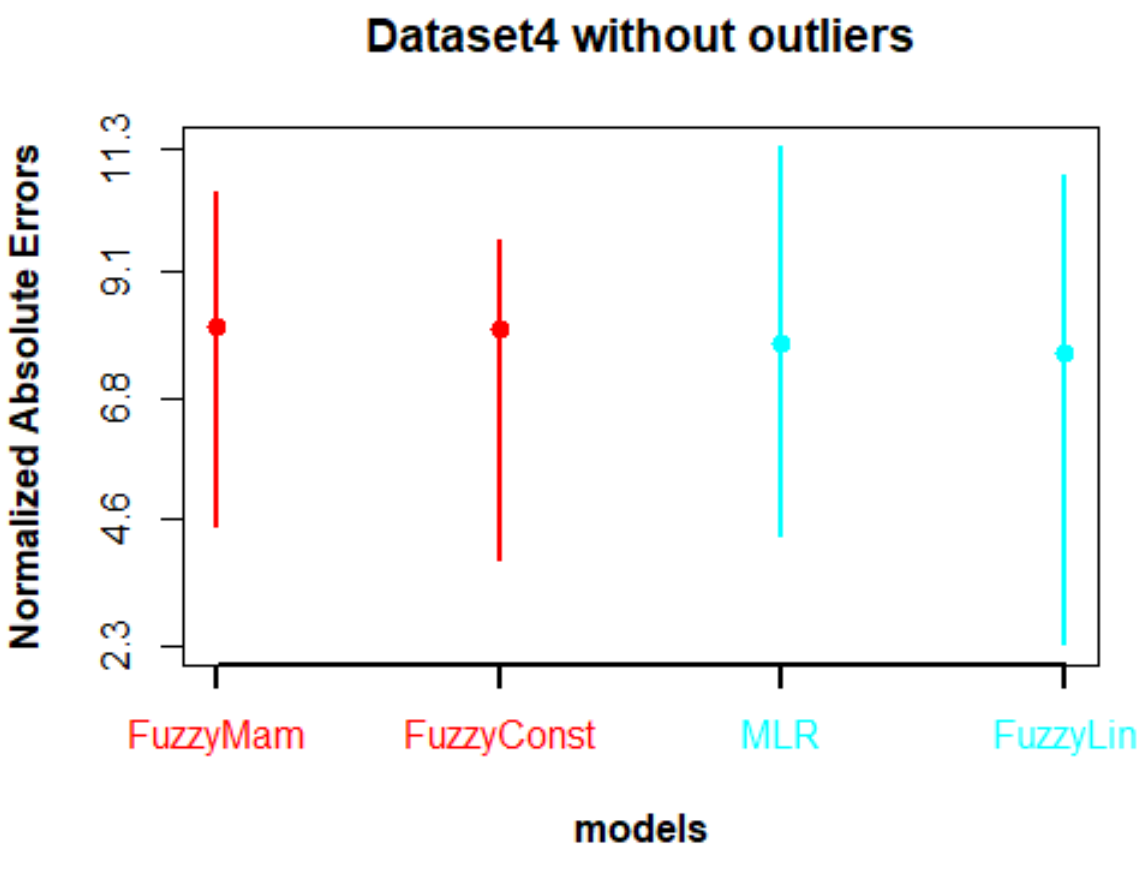

(d) Dataset 4

*Figure 5: Scott-Knott test results in datasets without outliers*

To visualize the effect of the outliers in the result of all models, a Scatterplot was extracted for the Sugeno Linear model in each dataset (with outliers and without outliers), where the x-axis is the actual effort and the y-axis is the estimated effort as shown in Figure 6. It is evident that removing the outliers decreased the drifting effect on the linear line generated. Note that Dataset 2 has no outliers.

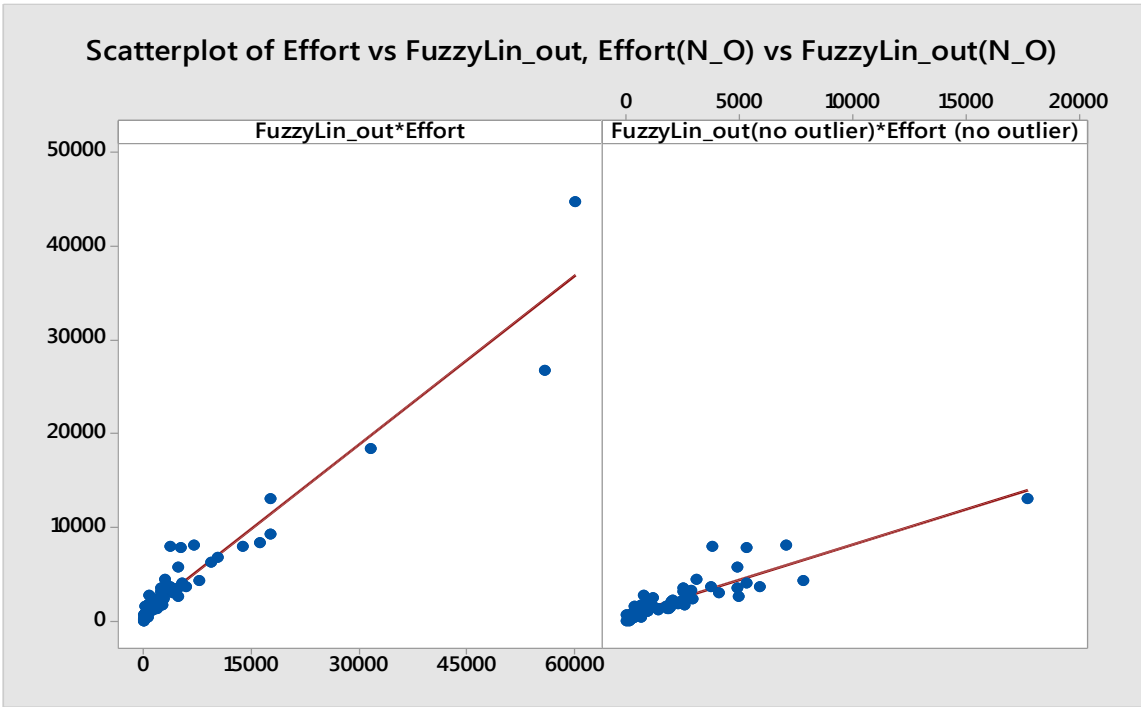

(a) Dataset 1

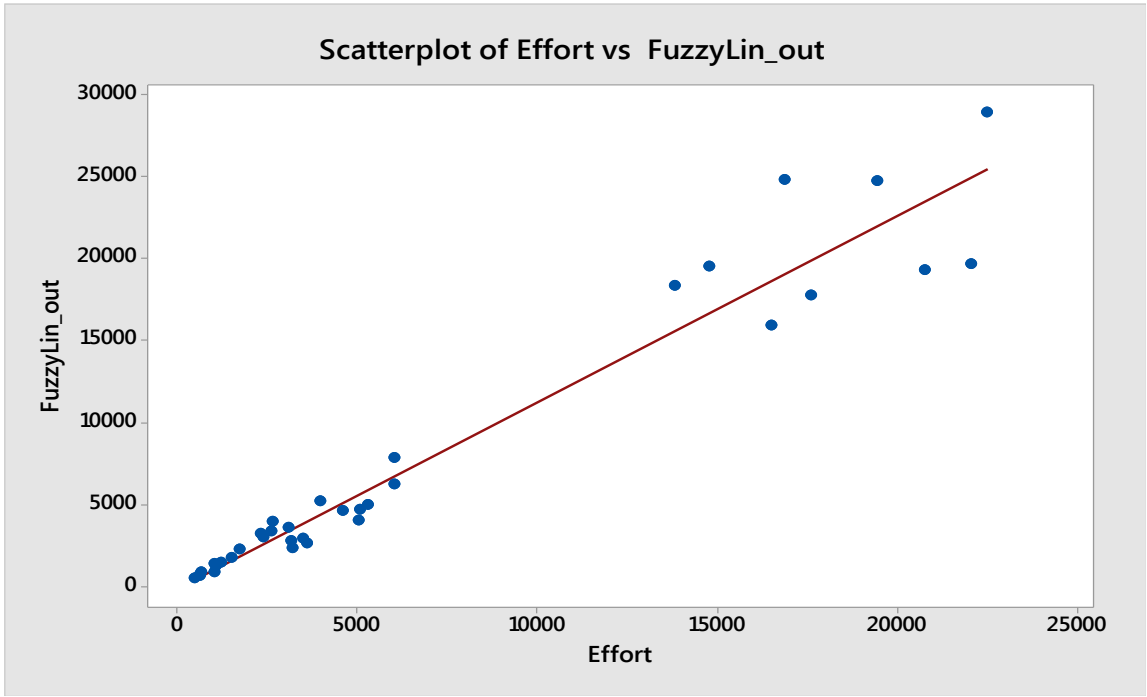

(b) Dataset 2

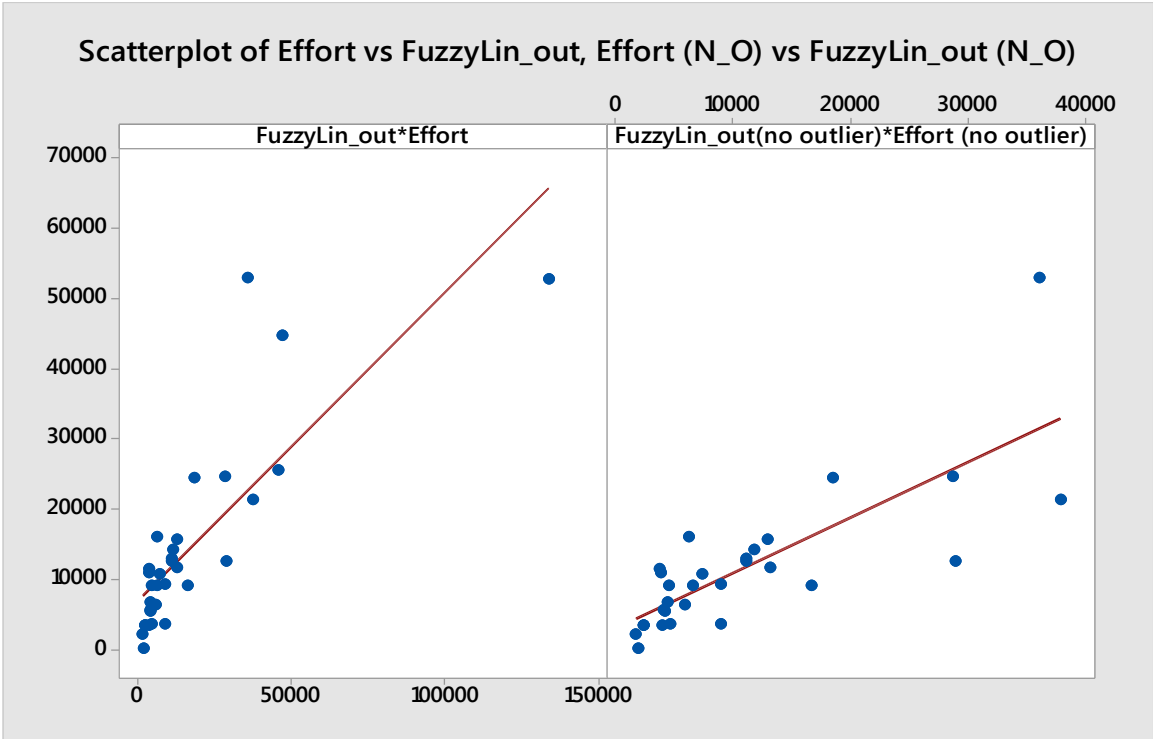

(c) Dataset 3

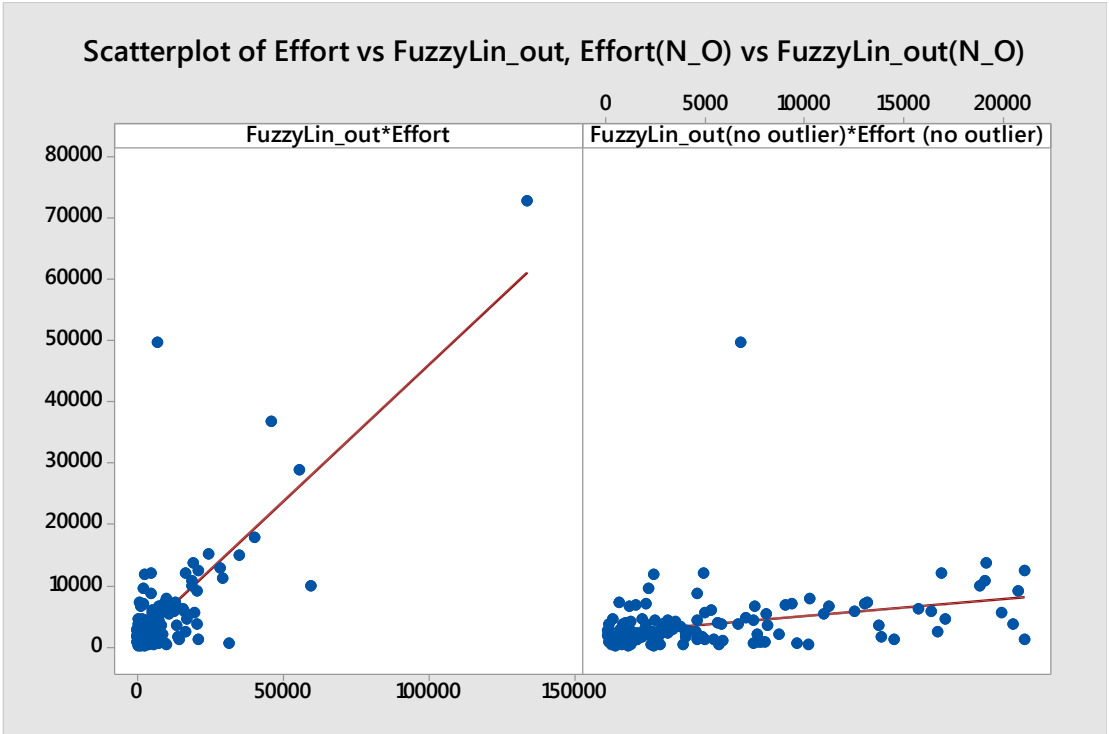

(d) Dataset 4

*Figure 6: Scatter plots for efforts predicted by FL Sugeno Linear and Actual effort with/without the presence of outliers.*

To validate the conclusion drawn about Sugeno Linear outperformance in estimating software costs, its results were compared to Forward Feed Artificial Neural Network model. The ANN model created were trained and tested in the 8 datasets that used in this research; 4 with outliers and 4 without outliers. A comparison between the MAE of both models is shown in Table 7. The Fuzzy Linear outperformed the ANN model in all the datasets.

*Table 7: Comparison between Sugeno FL and ANN model based on MAE*

| | ***With outliers*** | | | | ***Without outliers*** | | | |
|---|---|---|---|---|---|---|---|---|
| | ***Dataset 1*** | ***Dataset 2*** | ***Dataset 3*** | ***Dataset 4*** | ***Dataset 1*** | ***Dataset 2*** | ***Dataset 3*** | ***Dataset 4*** |
| ***Fuzzy Lin_out*** | 1842.61 | 1342.3 | 7241.36 | 4925.23 | 720.05 | 1342.92 | 4376.3 | 3613.67 |
| ***ANN_out*** | 2041.65 | 3208.2 | 8499.06 | 5694.96 | 961.8 | 3208.23 | 4399.3 | 4492.82 |

### *6.2 Answers to Research Questions*

RQ1: What is the impact of using regression analysis on tuning the parameters of fuzzy models?

Based on the results in Section 6, we conclude that Sugeno linear FL model combined the fuzziness characteristics of fuzzy logic models with the nature of regression models. The different membership functions and rules used, allowed the model to cope with software parameter complexity. The Sugeno linear FL model showed stable behavior and high accuracy compared to the MLR and other models as shown in Scott-Knott plots. We conclude that regression analysis can assist in designing fuzzy logic models, especially the parameters of Sugeno fuzzy with linear output.

RQ2: How might data heteroscedasticity affect the performance of such models?

A heteroscedasticity issue appears when the productivity (effort/size) fluctuates among projects in the same dataset. To see this impact, we divided the datasets into four sets containing different groups of productivity as described in Section 4. Heteroscedasticity appeared in the third dataset. Multiple tests were applied on all the datasets to identify the difference in performance. We concluded that heteroscedasticity had a detrimental effect on the performance of fuzzy logic models, but when we applied statistical tests, we found that in those datasets where heteroscedasticity existed, none of the models were statistically different. However, we concluded that the Sugeno linear FL model outperformed other models in the presence and absence of the heteroscedasticity issue.

RQ3: How do outliers affect the performance of the models?

After generating four datasets, we extracted the outliers from each testing dataset. We then applied the same error measurements and statistical tests on each, as described in Section 6.2. We extracted interval plots for mean absolute error of predicted results with and without outliers as shown in Figure 4. A general improvement was noticed after removing outliers, since we observed a major decrease in MAE and the interval range shortened (decreased). Furthermore, results showed that datasets became more homogenous after removing the outliers. We also found that the models tend to underestimate in the presence of outliers and overestimate when outliers are removed, yet the performance of all models improved when outliers were removed. Despite the fact that outliers affect the performance of the models, the Sugeno linear model still proved to be the best performing model.

We have proven in this research that the Sugeno linear fuzzy logic model outperforms other models in the presence of outliers, absence of outliers and when the dataset is homogenous or heterogeneous. When we mentioned "the same model for all projects was therefore not practical", this is because each model was trained using a different dataset. To predict the effort of a new project in a certain organization, the Sugeno linear fuzzy logic model can be re-trained on some historical projects in the same organization and thus, can be used to predict future projects

## 7. Threats to Validity

This section presents threats to the validity of this research, specifically internal and external validity. Regarding internal validity, the datasets used in this research work were divided randomly into training and testing groups, 70% and 30%, respectively. Although the leave-one-out (LOO) cross validation method is less biased than the random splitting method [56], the technique was not implemented because of the difficulty of designing fuzzy logic models with the LOO method. In order to apply the LOO in our work, more than 1,000 models would have had to be manually generated in order to conduct all experiments with and without outliers, which is extremely difficult to implement. In our case, fuzzy logic models were designed manually from the training datasets.

External validity questions whether or not the findings can be generalized. In this work four datasets were generated from the ISBSG dataset with projects ranked A and B. Moreover, unbiased performance evaluation criteria and statistical tests were used to affirm the validity of the results. So, we can conclude that the results of this paper can be generalized to a large degree. However, using more datasets would yield more robust results.

# 8. Conclusions

This paper compared four models: Sugeno linear FL, Sugeno constant FL, Mamdani FL and MLR. Models were trained and tested using four datasets extracted from ISBSG. Then, the performance of the models was analyzed by applying various unbiased performance evaluation criteria and statistical tests that included: MAE, MBRE, MIBRE, SA and Scott-Knott. Then, outliers were removed and the same tests were repeated in order to draw a conclusion about superior models. The inputs for all models were software size (AFP), team size and resource level, while the output was software effort. Three main questions were posed at the beginning of the research:

RQ1: What is the impact of using regression analysis on tuning the parameters of fuzzy models?

RQ2: How might data heteroscedasticity affect the performance of such models?

RQ3: How do outliers affect the performance of the models?

Based on the discussions of the results in Section 6, we conclude the following:

1) Combining the multiple linear regression concept with the fuzzy concept, especially in the Sugeno fuzzy model with linear output, led to a better design of fuzzy models, especially by learning the optimized number of model inputs, as well as the parameters for the fuzzy linear model.

2) Where a heteroscedasticity problem exists, the Sugeno fuzzy model with linear output was the best performing among all models. However, we note that although the Sugeno linear is the superior model, it is not statistically different from the others.

3) When outliers were removed, the performance of all the models improved. The Sugeno fuzzy model with linear output did however remain the superior model.

In conclusion, results showed that the Sugeno fuzzy model with linear output outperforms Mamdani and Sugeno with constant output. Furthermore, Sugeno with linear output was found to be statistically different than the other models on most of the datasets using Wilcoxon statistical tests in the absence of the heteroscedasticity problem. The validity of the results was also confirmed using the Scott-Knott test. Moreover, results showed that despite heteroscedasticity and the influence of outliers on the performance of all the fuzzy logic models, the Sugeno fuzzy model with linear output remained the model with the best performance.

# Acknowledgement

The authors thank part-time research assistant Omnia Abu Waraga, Eng., for conducting experiments for this paper.

Ali Bou Nassif extends thanks to the University of Sharjah for supporting this research through the Seed Research Project number 1602040221-P. The research was also supported by the Open UAE Research and Development Group at the University of Sharjah.

Mohammad Azzeh is grateful to the Applied Science Private University, Amman, Jordan, for the financial support granted to conduct this research.

# References

[1] M. Jorgensen and M. Shepperd, "A systematic review of software development cost estimation studies," *IEEE Trans. Softw. Eng.*, vol. 33, no. 1, 2007.

[2] F. J. Heemstra, "Software cost estimation," *Inf. Softw. Technol.*, vol. 34, no. 10, pp. 627–639, 1992.

[3] M. Azzeh, A. B. Nassif, and S. Banitaan, "Comparative analysis of soft computing techniques for predicting software effort based use case points," *IET Softw.*, vol. 12, no. 1, pp. 19–29, 2018.

[4] R. Silhavy, P. Silhavy, and Z. Prokopova, "Analysis and selection of a regression model for the Use Case Points method using a stepwise approach," *J. Syst. Softw.*, vol. 125, 2017.

[5] R. Silhavy, P. Silhavy, and Z. Prokopova, "Evaluating subset selection methods for use case points estimation," *Inf. Softw. Technol.*, vol. 97, pp. 1–9, May 2018.

[6] C. Lopez-Martin, C. Yáñez-Márquez, and A. Gutierrez-Tornes, "A fuzzy logic model for software development effort estimation at personal level," *MICAI 2006 Adv. Artif. Intell.*, pp. 122–133, 2006.

[7] L. A. Zadeh, "Fuzzy sets," *Inf. Control*, vol. 8, no. 3, pp. 338–353, 1965.

[8] A. Idri and A. Abran, "COCOMO Cost Model Using Fuzzy Logic," in *7th International Conference on Fuzzy Theory and Technology*, 2000, pp. 1–4.

[9] C. López-Martin, C. Yáñez-Márquez, and A. Gutiérrez-Tornés, "Predictive accuracy comparison of fuzzy models for software development effort of small programs," *J. Syst. Softw.*, vol. 81, no. 6, pp. 949–960, 2008.

[10] M. Hosni, A. Idri, A. Abran, and A. B. Nassif, "On the value of parameter tuning in heterogeneous ensembles effort estimation," *Soft Comput.*, vol. 22, no. 18, pp. 5977–6010, 2018.

[11] N. Mittas and L. Angelis, "Ranking and clustering software cost estimation models through a multiple comparisons algorithm," *IEEE Trans. Softw. Eng.*, vol. 39, no. 4, 2013.

[12] M. Shepperd and S. MacDonell, "Evaluating prediction systems in software project estimation," *Inf. Softw. Technol.*, vol. 54, no. 8, pp. 820–827, 2012.

[13] T. Foss, E. Stensrud, B. Kitchenham, and I. Myrtveit, "A Simulation Study of the Model Evaluation Criterion MMRE," *IEEE Trans. Softw. Eng.*, vol. 29, no. 11, no. 11, pp. 985–995, 2003.

[14] A. Idri, I. Abnane, and A. Abran, "Evaluating Pred(p) and standardized accuracy criteria in software development effort estimation," *J. Softw. Evol. Process*, vol. 30, no. 4, 2017.

[15] I. Myrtveit and E. Stensrud, "Validity and reliability of evaluation procedures in comparative studies of effort prediction models," *Empir. Softw. Eng.*, vol. 17, no. 1, pp. 23–33, 2012.

[16] ISBSG, "International Software Benchmarking Standards Group." .

[17] H. Liu, J. Wang, Y. He, and R. A. R. Ashfaq, "Extreme learning machine with fuzzy input and fuzzy output for fuzzy regression," *Neural Comput. Appl.*, vol. 28, no. 11, pp. 3465–3476, Nov. 2017.

[18] A. R. Gray and S. G. MacDonell, "A comparison of techniques for developing predictive models of software metrics," *Inf. Softw. Technol.*, vol. 39, no. 6, pp. 425–437, 1997.

[19] N. Garcia-Diaz, C. Lopez-Martin, and A. Chavoya, "A comparative study of two fuzzy logic models for software development effort estimation," *Procedia Technol.*, vol. 7, pp. 305–314, 2013.

[20] Z. Xu and T. M. Khoshgoftaar, "Identification of fuzzy models of software cost estimation," *Fuzzy Sets Syst.*, vol. 145, no. 1, pp. 141–163, 2004.

[21] M. A. Ahmed, M. O. Saliu, and J. AlGhamdi, "Adaptive fuzzy logic-based framework for software development effort prediction," *Inf. Softw. Technol.*, vol. 47, no. 1, pp. 31–48, 2005.

[22] C. L. Martin, J. L. Pasquier, C. M. Yanez, and A. G. Tornes, "Software development effort estimation using

fuzzy logic: a case study," in *Computer Science, 2005. ENC 2005. Sixth Mexican International Conference on*, 2005, pp. 113–120.

[23] A. Sheta, "Software effort estimation and stock market prediction using takagi-sugeno fuzzy models," in *Fuzzy Systems, 2006 IEEE International Conference on*, 2006, pp. 171–178.

[24] I. Attarzadeh and S. H. Ow, "Software development effort estimation based on a new fuzzy logic model," *Int. J. Comput. Theory Eng.*, vol. 1, no. 4, p. 473, 2009.

[25] C. López-Martín and A. Abran, "Neural networks for predicting the duration of new software projects," *J. Syst. Softw.*, vol. 101, 2015.

[26] H. K. Verma and V. Sharma, "Handling imprecision in inputs using fuzzy logic to predict effort in software development," in *Advance Computing Conference (IACC), 2010 IEEE 2nd International*, 2010, pp. 436–442.

[27] A. B. Nassif, L. F. Capretz, and D. Ho, "Estimating software effort based on use case point model using Sugeno Fuzzy Inference System," in *Proceedings - International Conference on Tools with Artificial Intelligence, ICTAI*, 2011, pp. 393–398.

[28] A. B. Nassif, L. F. Capretz, and D. Ho, "A Regression Model with Mamdani Fuzzy Inference System for Early Software Effort Estimation Based on Use Case Diagrams," in *Third International Conference on Intelligent Computing and Intelligent Systems*, 2011, vol. Guangzhou, pp. 615–620.

[29] I. Attarzadeh and S. H. Ow, "Improving estimation accuracy of the COCOMO II using an adaptive fuzzy logic model," in *Fuzzy Systems (FUZZ), 2011 IEEE International Conference on*, 2011, pp. 2458–2464.

[30] C. Lopez-Martin, "A fuzzy logic model for predicting the development effort of short scale programs based upon two independent variables," *Appl. Soft Comput.*, vol. 11, no. 1, pp. 724–732, 2011.

[31] S. Kumar and V. Chopra, "Neural network and fuzzy logic based framework for software development effort estimation," *Int. J. Adv. Res. Comput. Sci. Softw. Eng.*, vol. 3, no. 5, 2013.

[32] X. Huang, L. F. Capretz, J. Ren, and D. Ho, "A neuro-fuzzy model for software cost estimation," in *Quality Software, 2003. Proceedings. Third International Conference on*, 2003, pp. 126–133.

[33] X. Huang, D. Ho, J. Ren, and L. F. Capretz, "Improving the COCOMO model using a neuro-fuzzy approach," *Appl. Soft Comput.*, vol. 7, no. 1, pp. 29–40, 2007.

[34] S.-J. Huang and N.-H. Chiu, "Applying fuzzy neural network to estimate software development effort," *Appl. Intell.*, vol. 30, no. 2, pp. 73–83, 2009.

[35] J. Wong, D. Ho, and L. F. Capretz, "An investigation of using neuro-fuzzy with software size estimation," in *Software Quality, 2009. WOSQ'09. ICSE Workshop on*, 2009, pp. 51–58.

[36] U. R. Saxena and S. P. Singh, "Software effort estimation using neuro-fuzzy approach," in *Software Engineering (CONSEG), 2012 CSI Sixth International Conference on*, 2012, pp. 1–6.

[37] W. L. Du, L. F. Capretz, A. B. Nassif, and D. Ho, "A Hybrid Intelligent Model for Software Cost Estimation," *J. Comput. Sci.*, vol. 9, no. 11, pp. 1506–1513, 2013.

[38] A. B. Nassif, "Software Size and Effort Estimation from Use Case Diagrams Using Regression and Soft Computing Models," University of Western Ontario, 2012.

[39] A. B. Nassif, M. Azzeh, L. F. Capretz, and D. Ho, "Neural network models for software development effort estimation: a comparative study," *Neural Comput. Appl.*, vol. 27, no. 8, pp. 2369–2381, 2016.

[40] E. Manalif, L. F. Capretz, A. B. Nassif, and D. Ho, "Fuzzy-ExCOM software project risk assessment," in *Proceedings - 2012 11th International Conference on Machine Learning and Applications, ICMLA 2012*, 2012, vol. 2, pp. 320–325.

[41] E. Ehsani, N. Kazemi, E. U. Olugu, E. H. Grosse, and K. Schwindl, "Applying fuzzy multi-objective linear programming to a project management decision with nonlinear fuzzy membership functions," *Neural Comput. Appl.*, vol. 28, no. 8, pp. 2193–2206, Aug. 2017.

[42] E. H. Mamdani, "Application of Fuzzy Logic to Approximate Reasoning Using Linguistic Synthesis,"

*Comput. IEEE Trans.*, vol. C-26, no. 12, pp. 1182–1191, 1977.

[43] M. Sugeno and T. Yasukawa, “A fuzzy-logic-based approach to qualitative modeling,” *Fuzzy Syst. IEEE Trans.*, vol. 1, no. 1, pp. 7–31, 1993.

[44] A. Mittal, K. Parkash, and H. Mittal, “Software cost estimation using fuzzy logic,” *ACM SIGSOFT Softw. Eng. Notes*, vol. 35, no. 1, no. 1, pp. 1–7, 2010.

[45] S. Sotirov *et al.*, “Application of the Intuitionistic Fuzzy InterCriteria Analysis Method with Triples to a Neural Network Preprocessing Procedure,” *Comput. Intell. Neurosci.*, vol. 2017, pp. 1–9, 2017.

[46] C. Chi-Chung and L. Yi-Ting, “Enhanced Ant Colony Optimization with Dynamic Mutation and Ad Hoc Initialization for Improving the Design of TSK-Type Fuzzy System,” *Comput. Intell. Neurosci.*, vol. 2018, pp. 1–16, 2018.

[47] M. Negnevitsky, *Artificial Intelligence: A Guide to Intelligent Systems*. Addison Wesley/Pearson, 2011.

[48] S. Chatterjee and A. S. Hadi, *Regression analysis by example*. John Wiley & Sons, 2015.

[49] M. Azzeh, A. B. Nassif, S. Banitaan, and F. Almasalha, “Pareto efficient multi-objective optimization for local tuning of analogy-based estimation,” *Neural Comput. Appl.*, vol. 27, no. 8, pp. 2241–2265, 2016.

[50] L. L. Minku and X. Yao, “How to Make Best Use of Cross-company Data in Software Effort Estimation?,” in *Proceedings of the 36th International Conference on Software Engineering*, 2014, pp. 446–456.

[51] S. Kopczyńska, J. Nawrocki, and M. Ochodek, “An empirical study on catalog of non-functional requirement templates: Usefulness and maintenance issues,” *Inf. Softw. Technol.*, vol. 103, pp. 75–91, Nov. 2018.

[52] V. Cheng, C.-H. Li, J. T. Kwok, and C.-K. Li, “Dissimilarity learning for nominal data,” *Pattern Recognit.*, vol. 37, no. 7, pp. 1471–1477, 2004.

[53] A. J. Scott and M. Knott, “A cluster analysis method for grouping means in the analysis of variance,” *Biometrics*, pp. 507–512, 1974.

[54] M. Azzeh and A. B. Nassif, “Analyzing the relationship between project productivity and environment factors in the use case points method,” *J. Softw. Evol. Process,* 2017.

[55] J. Han, M. Kamber, and J. Pei, *Data Mining: Concepts and Techniques*. Morgan Kaufmann, 2012.

[56] E. Kocaguneli and T. Menzies, “Software effort models should be assessed via leave-one-out validation,” *J. Syst. Softw.*, vol. 86, no. 7, pp. 1879–1890, Jul. 2013.